# KINEMATIC CLUES TO BAR EVOLUTION FOR GALAXIES IN THE LOCAL UNIVERSE: WHY THE FASTEST ROTATING BARS ARE ROTATING MOST SLOWLY.

Rotational kinematics of bars

J. FONT[1,2], J. E. BECKMAN[1,2,3], I. MARTÍNEZ-VALPUESTA[1,2], A. S. BORLAFF[1,2], P. A. JAMES[4], S. DÍAZ-GARCÍA[5], B. GARCÍA-LORENZO[1,2], A. CAMPS-FARIÑA[1,2], L. GUTIÉRREZ[6], P. AMRAM[7]

1. Instituto de Astrofísica de Canarias, c/ Vía Láctea, s/n, E38205, La Laguna, Tenerife, Spain: jfont@iac.es, jeb@iac.es, imv@iac.es, asborlaff@iac.es, artemic@iac.es
2. Departamento de Astrofísica. Universidad de La Laguna, Tenerife, Spain.
3. Consejo Superior de Investigaciones Científicas, Spain.
4. Astrophysics Research Institute, Liverpool John Moores University, IC2, Liverpool Science Park, 146 Brownlow Hill, Liverpool L3 5RF, UK: P.A.James@ljmu.ac.uk
5. Astronomy Research Group, University of Oulu, FI-90014 Finland: simon.diazgarcia@oulu.fi
6. Universidad Nacional Autónoma de México, Instituto de Astronomía, Ensenada, B. C., Mexico: leonel@astrosen.unam.mx
7. Laboratoire d'Astrophysique de Marseille, Université d'Aix-Marseille & CNRS, UMR7326, 38 rue F. Juliot-Curie, 13388 Marseille Cedex 13, France: philippe.amram@oamp.fr

## ABSTRACT

We have used Spitzer images of a sample of 68 barred spiral galaxies in the local universe to make systematic measurements of bar length and bar strength. We combine these with precise determinations of the corotation radii associated with the bars, taken from our previous study which used the phase change from radial inflow to radial outflow of gas at corotation, based on high resolution two-dimensional velocity fields in Hα taken with a Fabry-Pérot spectrometer. After presenting the histograms of the derived bar parameters, we study their dependence on the galaxy morphological type and on the total stellar mass of the host galaxy, and then produce a set of parametric plots. These include the bar pattern speed versus bar length, the pattern speed normalized with the characteristic pattern speed of the outer disk versus the bar strength, and the normalized pattern speed versus $\mathcal{R}$, the ratio of corotation radius to bar length. To provide guide-lines for our interpretation we used a recently published simulations, including disk and dark matter halo components. Our most striking conclusion is that bars with values of $\mathcal{R} < 1.4$, previously considered dynamically fast rotators, can be among the slowest rotators both in absolute terms and when their pattern speeds are normalized. The simulations confirm that this is because as the bars are braked they can grow longer more quickly than the outward drift of the corotation radius. We conclude that dark matter halos have indeed slowed down the rotation of bars on Gyr timescales.

*Subject headings*: galaxies: spirals - galaxies: fundamental parameters - galaxies: kinematics and dynamics - galaxies: structure – galaxies: evolution - techniques: interferometric.

1. INTRODUCTION

In spite of the early classification of spiral galaxies into "normal" and "barred" spirals, as observations have accumulated it has become clear that the description "normal" could better have been given to the barred galaxies, as these form the majority of the spirals. Almost two thirds of galaxies in the local universe have bars (Knapen et al. 2000; Laurikainen et al. 2004; Menéndez-Delmestre et al. 2007), of which close to one half are strongly barred (Sellwood & Wilkinson 1993). In order to understand the evolution of disk galaxies in general it is therefore of particular interest to study the evolution of bars. They interact dynamically with the other structural components, such as the disks they inhabit, the bulges (if any) and the dark matter halos. This interaction, in which angular momentum interchange is important, makes bars useful signposts to evolution in general. In the last 15 years particular attention has been drawn to the use of the angular rotation rates of bars as tests of the evolutionary braking brought about by dark matter halos. In a seminal article by Debattista & Sellwood (2000), they performed a set of simulations showing the evolution of bars in this context. In this study they took the ratio, $\mathcal{R}$, of the corotation radius to the bar length as an index of whether a bar could be characterized as fast or slow. Starting from the theoretical argument by Contopoulos (1980) showing that $\mathcal{R}$ should be close to, and a little greater than, unity they proposed that bars in the range $1 < \mathcal{R} < 1.4$, (i.e. where corotation is not far from the end of the bar) should be considered fast, because dynamical braking would move corotation progressively further out compared to the bar length. In that article they noted that to use this as a test for dynamical braking was, at that time, made difficult by the rather small numbers of measurements of the pattern speed, and pointed to the Tremaine-Weinberg (Tremaine & Wienberg 1984) method as a preferred option for this. The few measured galaxies (Merrifield and Kuijken 1995; Gerssen et al. 1999) with published data using this technique had values of $R$ between 1.4 and 1. A number of less direct methods, using high spatial resolution two-dimensional gas dynamics, (e.g. Lindblad et al. 1996), as well as arguments using the shapes of dust lanes (van Albada & Sanders 1982; Athanassoula, 1992) also gave values in this range.

In the intervening years significant effort has been devoted to measurements designed to give reliable values of the corotation radius to the bar length ratio. For a review of these used until 2008 we refer the reader to Rautiainen et al. (2008). Here we list some of the most representative methods, some of which measure the pattern speed of the bar while others determine the corotation radius directly or indirectly. They are: (*i*) the well-known Tremaine-Winberg method (Tremaine & Weinberg 1984), and its later modification to include multiple density wave patterns (Meidt et al. 2008a, 2008b) combines photometric data with velocities derived from long slit spectra placed parallel to the galaxy major axis (Kent 1987; Merrifield & Kruijken 1995; Gerssen et al. 1999; Debattista et al. 2002; Aguerri et al. 2003; Corsini et al. 2003; Debattista & Williams 2004; Zimmer et al. 2004; Rand & Wallin 2004; Hernandez et al. 2005; Fathi et al. 2009; Corsini 2011, and references therein, Aguerri et al. 2015). (*ii*) using the gravitational potential distribution to reproduce the velocity map (Sanders & Tubbs 1980; England et al. 1990; Garcia-Burillo et al. 1993; Piñol-Ferrer et al. 2014). (*iii*) studying the morphologies of the non-circular velocity map of the gas in the galaxy (Canzian 1993; Sempere et al. 1995; Rand 1995; Canzian & Allen 1997; Font et al. 2011, 2014a,b). (*iv*) using numerical simulations of barred galaxies (Lindblad et al. 1996; Laine & Heller 1999; Weiner et al. 2001; Aguerri et al 2001; Perez et al. 2004; Rautiainen et al. 2005, 2008; Treuthardt et al. 2008)

A quantitative step forward in terms of the number of galaxies measured was made by Font & Beckman (Font et al. 2011, 2014a, b) who took full advantage of the high spectral and spatial resolution made possible with scanning Fabry-Pérot spectroscopic technology to analyze the two-dimensional velocity fields of over 100 galaxies observed in H$\alpha$. Rather than trying to determine the pattern speeds of the structures in these galaxies, they directed their attention directly to measuring the corotation radius (to be precise the corotation radii, as they found radii associated with more than one structural component of the disk in virtually all cases), they used the property of particles undergoing streaming motions in a barred potential: that their non-circular components of velocity should exhibit a phase change at the corotation radius. (Kalnajs 1978; Miller & Smith 1979; Sparke & Sellwood 1987), from inward motion to outward motion as the corotation radius is crossed. Its short response timescale makes the gas a sensitive detector of this phase change, and the resulting values of the corotation radius are in most cases sharply defined. We use these observations here as a key step towards our determination of $\mathcal{R}$ in the galaxies selected for analysis in the present article.

But the value of $\mathcal{R}$ does not depend only on the corotation radius, it also requires a well determined value of the bar length. Although this measurement is in principle less complicated than that of corotation, it suffers from the fact that there is no consensus in the astronomical community in defining a magnitude that measures the length of a bar, not only for observed galaxies but also for simulations. This is manifested in a diversity of methods developed to measure the size of the bar, each of them responding to a different concept of where to place the radius of its end (in Athanassoula & Misiriotis (2002), up to eight different methods to calculate the bar length are reported). We can distinguish three principal different techniques that have been applied to calculate the bar length: (*i*) Ellipse fitting to the isophotes of the galaxy; this standard method assumes that the end of the bar is found at the radius where the ellipticity of the ellipses reaches its maximum. It was initially applied to real galaxies in Wozniak & Pierce (1991) and in Wozniak et al. (1995) based on the technique described by Jedrzejewski (1987). This method has been extensively used on real galaxies (Laine et al. 2002; Sheth et al. 2003; Erwin & Sparke 2003; Erwin 2004, 2005; Gadotti & Souza 2006; Gadotti et al. 2007; Marinova & Jogee 2007; Aguerri et al. 2009; Díaz-García et al. 2016). A variation of this technique has also been used to determine the bar length in simulated galaxies (Athanassoula et al., 1990; Villa-Vargas et al. 2009, 2010; Athanassoula 2014). (*ii*) Various different measurements based on the Fourier decomposition of the galaxy image (Ohta et al. 1990; Quillen et al. 1994; Debattista & Sellwood 2000; Aguerri et al. 2001, 2003, 2005; Athanassoula & Misiriotis 2002; Laurikainen et al. 2009; Athanassoula, 2014). (*iii*) Visual estimation of the bar size from images (Kormendy 1979; Martin 1995). This is the most direct method but it is very dependent on the quality of the images analyzed, so it should be used only as a first approximation method; alternatively it can be efficiently used as a validation for the results obtained when any of the former method is applied.

The bar strength is another parameter used to characterize bars for which (as with the bar length) there is no agreed definition. In consequence there is a considerable variety of methods to calculate this parameter found in the literature: (*i*) using just the ellipticities, which can be calculated either by ellipse fitting or measuring the minor axis of the bar (Martin, 1995; Martinet & Friedli 1997; Aguerri 1999; Knapen et al. 2000; Laine et al., 2002). (*ii*) from the Fourier decompositions of the galaxy image (Ohta et al. 1990; Marquez et al. 1996; Laurikainen & Salo 2002; Laurikainen et al. 2002; Athanassoula 2003; Laurikainen et al. 2005; Athanassoula et al. 2012; Díaz-García et al. 2016; Martínez-Valpuesta et al. 2017). (*iii*) from the torque of the bar measured from the gravitational potential inferred from the infrared image (Combes & Sanders 1981; Quillen et al. 1994; Buta & Block 2001; Berentzen et al. 2007; Tiret & Combes 2008; Salo et al. 2010; Díaz-García et al. 2016).

In this article we will, in section 2 describe briefly the observational data used for both corotation radius and bar length measurements, and the methods of data analysis employed. The analysis is not confined to deriving the corotation radius and the bar length; we also compute the bar strength for all the objects under study. In section 2 we also define and determine a characteristic angular speed for the disk, defined by dividing the asymptotic rotational velocity by the radius at $r_{25}$, which will later be used to scale our pattern speeds. In section 3 we give our results, which include the statistics of the basic bar parameters, their variation with morphological type, their dependence on galaxy mass, and the relationships between them. In section 4 we use specific simulations as broad guide-lines to see whether our results can be explained in terms of evolutionary sequences for bar kinematics. In section 5 we question the by now conventional definition of the range of $\mathcal{R}$ which defines a fast bar, and draw new implications for the phenomenon of the braking of bar rotation by dark matter halos. Finally in section 6 we give our conclusions.

## 2. OBSERVATIONAL DATA AND DATA ANALYSIS

In Font et al. (2014a) a total of 104 nearby galaxies observed with a Fabry-Pérot interferometer were studied. A detailed description of the sample selection criteria is given in Font et al. (2014a). From that sample we have selected those galaxies which are classified as barred, or show a bar-like central structure, and also have a spiral arm structure, obtaining a subset of 68 barred spiral galaxies, which are analyzed in the present study.

### 2.1. The Observational Data

To perform the measurements of the bar size and bar strength of each galaxy, we used images from different surveys according to their availability and quality (Table 1 lists the objects of the present study, and also includes in column 3 the survey from which the images are obtained); to minimize dust effects on the morphology we gave priority to images in the infrared, so most of the images in our sample are 3.6 μm infrared images taken from the Spitzer archive[1], which offers the best infrared images publicly available online. We also found infrared images in the J, H and Ks bands in the 2MASS survey[2] for only five galaxies of our sample, but with a quality not sufficient to make the calculations, except for the galaxy UGC2855 for which we used its image in the J band. For those galaxies for which no infrared image is available, we turned to images from the Sloan Digital Sky Survey in the r band (data release 12 of the SDSS III[3]). Finally, the images of only three galaxies of our sample (UGC2080, UGC11124 and UGC12276) are taken from the ESO Digitized Sky Survey[4] (DSS2-red), as this is the only survey that provides data with enough resolution to make reliable calculations for these objects.

---

[1] http://sha.ipac.caltech.edu/applications/Spitzer/SHA/

[2] https://irsa.ipac.caltech.edu/applications/2MASS/PubGalPS/

The kinematical information is obtained from the high resolution velocity fields of the ionized gas, which are extracted from a Fabry-Pérot data cube. The Fabry-Pérot interferometer maps the Hα emission line across the whole field covering the observed galaxy, and produces a [$x,y,\lambda'$] or [$x,y,v_{l.o.s.}$] 3D datacube (where $x$, $y$ are the two spatial coordinates, $\lambda'$ and $v_{l.o.s.}$ are the Hα redshifted wavelength and the velocity in the line of sight of the ionized gas, respectively) with high spectral and angular resolutions, after performing the phase calibration and the wavelength calibration.

The majority of velocity maps of the galaxies of this sample, 64 of a total set of 68, are taken from the GHASP survey (Gassendi HAlpha survey of SPirals[5], Epinat et al. 2008). This survey consist of a large sample of 203 spiral and irregular galaxies observed with a Fabry-Pérot instrument at the 1.93m telescope of the Observatoire de Haute de Provence, in France, during the period 1998-2004. The data cubes with a pixel scale of 0.68 arcsec/pix obtained with this interferometer have an angular resolution limited by the seeing value with an averaged value of ~ 3arcsec, and the spectral resolution is ~ 16km·s$^{-1}$ in Hα. The remaining four galaxies (UGC3013, UGC5303, UGC6118 and UGC7420) were observed with GHαFaS (Galaxy Hα Fabry-Pérot System, Hernandez et al. 2008). The observations were carried out in several runs at the William Herschel Telescope, Roque de los Muchachos Observatory, La Palma, Spain, in the period between 2010 and 2014. This instrument has a field of view of 3.4 arcmin$^2$ and gives data cubes with a spectral resolution of ~ 8 km·s$^{-1}$ in Hα, and with an average seeing limited angular resolution of ~ 1.2arcsec.

**Table 1.**
Properties of the galaxies

| Object Name | | Morphology | | Survey Image | D (Mpc) | $r_{25}$ (arcsec) | $i$ (º) | PA (º) | $V_{asym}$ (km·s$^{-1}$) | $M_{stellar}$ ($M_{sun}$) |
|---|---|---|---|---|---|---|---|---|---|---|
| UGC | NGC | | | | | | | | | |
| (1) | (2) | (3) | (4) | (5) | (6) | (7) | (8) | (9) | (10) | (11) |
| 508 | 266 | SB(rs)ab | | Spitzer | 63.8 | 88.55 | 25 | 123 | 553 | 8.74·10^10 |
| 763 | 428 | SAB(s)m | SAB(s)dm | Spitzer | 12.7 | 122.2 | 54 | 117 | 104 | 2.06·10^9 |
| 1256 | 672 | SB(s)cd | (R')SB(s)d | Spitzer | 7.2 | 217.35 | 76 | 76 | 85 | 1.06·10^9 |
| 1317 | 674 | SAB(r)c | | Spitzer | 42.2 | 134 | 73 | 73 | 205 | 1.23·10^11 |
| 1437 | 753 | SAB(rs)bc | | Spitzer | 66.8 | 75.35 | 47 | 47 | 218 | 4.02·10^10 |
| 1736 | 864 | SAB(rs)c | SAB(s)bc | Spitzer | 17.6 | 140.3 | 35 | 35 | 193 | 1.04·10^10 |
| 1913 | 925 | SAB(s)d | | Spitzer | 9.3 | 314.15 | 48 | 48 | 105 | 1.74·10^10 |
| 2080 | IC239 | SAB(rs)cd | | DSS | 13.7 | 137.15 | 25 | 25 | 131 | 1.21·10^10 |
| 2855 | | SABc | | 2MASS | 17.5 | 130.95 | 68 | 68 | 229 | 6.28·10^10 |
| 3013 | 1530 | SB(rs)b | | Spitzer | 37.0 | 137.15 | 55 | 195 | 212 | 8.65·10^10 |
| 3463 | IC2166 | SAB(s)bc | | Spitzer | 38.6 | 90.6 | 63 | 110 | 168 | 1.72·10^11 |
| 3685 | | SB(rs)b | | Spitzer | 26.3 | 99.35 | 12 | 298 | 102 | 1.84·10^10 |
| 3709 | 2342 | S pec | | Spitzer | 70.7 | 41.4 | 55 | 232 | 241 | 1.02·10^11 |
| 3740 | 2276 | SAB(rs)c | | Spitzer | 17.1 | 84.55 | 48 | 247 | 87 | 9.29·10^10 |
| 3809 | 2336 | SAB(r)bc | | Spitzer | 32.9 | 212.4 | 58 | 357 | 258 | 3.46·10^11 |

---

[3] http://www.sdss.org/dr12/

[4] http://archive.eso.org/dss/dss

[5] http://cesam.lam.fr/fabryPérot/

|  |  |  |  |  |  |  |  |  |  |  |
|---|---|---|---|---|---|---|---|---|---|---|
| 4165 | 2500 | SB(rs)d | SAB(s)d | Spitzer | 11.0 | 86.5 | 41 | 265 | 80 | $9.42 \cdot 10^8$ |
| 4273 | 2543 | SB(s)b | SA$\underline{B}$(s)b | Spitzer | 35.4 | 70.35 | 60 | 212 | 200 | $2.65 \cdot 10^8$ |
| 4325 | 2552 | SA(s)m | (R')SAB(s)m | Spitzer | 10.9 | 104 | 63 | 57 | 85 | $3.77 \cdot 10^8$ |
| 4422 | 2595 | SAB(rs)c | | Spitzer | 58.1 | 94.85 | 25 | 36 | 353 | $6.76 \cdot 10^{10}$ |
| 4555 | 2649 | SAB(rs)bc | | Spitzer | 58.0 | 47.55 | 38 | 90 | 185 | $1.86 \cdot 10^{10}$ |
| 4936 | 2805 | SAB(rs)d | (R)SA(s)c pec | Spitzer | 25.6 | 189.3 | 13 | 294 | 230 | $4.36 \cdot 10^{10}$ |
| 5228 | | SB(s)c | (R$_2$')SA$\underline{B}$(s)bc | Spitzer | 24.7 | 73.65 | 72 | 120 | 125 | $3.35 \cdot 10^9$ |
| 5303 | 3041 | SAB(rs)c | SA(rs)$\underline{c}$ | Spitzer | 17.7 | 111.45 | 36 | 273 | 202 | $6.18 \cdot 10^9$ |
| 5319 | 3061 | (R')SB(rs)c | SAB(rs)b pec | Spitzer | 35.8 | 49.8 | 30 | 345 | 180 | $2.32 \cdot 10^{10}$ |
| 5510 | 3162 | SAB(rs)bc | SA(s)bc | Spitzer | 18.6 | 90.6 | 31 | 200 | 167 | $5.43 \cdot 10^9$ |
| 5532 | 3147 | SA(rs)bc | S$\underline{A}$B(rs)b | Spitzer | 41.1 | 116.7 | 32 | 147 | 398 | $1.50 \cdot 10^{10}$ |
| 5786 | 3310 | SAB(r)bc pec | SA(rs)bc pec | Spitzer | 14.2 | 92.7 | 53 | 153 | 80 | $2.55 \cdot 10^9$ |
| 5840 | 3344 | (R)SAB(r)bc | SAB(r)bc | Spitzer | 6.9 | 212.4 | 25 | 333 | 251 | $4.70 \cdot 10^9$ |
| 5842 | 3346 | SB(rs)cd | SB(rs)cd | Spitzer | 15.2 | 86.5 | 47 | 292 | 110 | $5.83 \cdot 10^9$ |
| 5982 | 3430 | SAB(rs)c | SAB(r)bc | Spitzer | 20.8 | 119.45 | 55 | 28 | 199 | $1.32 \cdot 10^{10}$ |
| 6118 | 3504 | (R)SAB(s)ab | (R$_1$')SAB(r,nl)a | Spitzer | 19.8 | 80.75 | 19 | 330 | 240 | $1.84 \cdot 10^{10}$ |
| 6537 | 3726 | SAB(r)c | SA$\underline{B}$(r)bc | Spitzer | 14.3 | 185 | 47 | 200 | 187 | $4.28 \cdot 10^9$ |
| 6778 | 3893 | SAB(rs)c | SA(s)c | Spitzer | 15.5 | 134 | 49 | 343 | 223 | $7.28 \cdot 10^9$ |
| 7021 | 4045 | SAB(r)a | (R$_1$'L)SAB(r,nl)ab | Spitzer | 26.8 | 80.75 | 56 | 266 | 175 | $1.98 \cdot 10^{10}$ |
| 7154 | 4145 | SAB(rs)d | SAB(rs)d | Spitzer | 16.2 | 176.65 | 65 | 275 | 145 | $5.96 \cdot 10^9$ |
| 7323 | 4242 | SAB(s)dm | (L)IAB(s)m | Spitzer | 8.1 | 150.35 | 51 | 38 | 84 | $1.50 \cdot 10^9$ |
| 7420 | 4303 | SAB(rs)bc | SAB(rs,nl)c | Spitzer | 20.0 | 193.7 | 29 | 135 | 177 | $2.09 \cdot 10^{10}$ |
| 7766 | 4559 | SAB(rs)cd | SB(s)cd | Spitzer | 13.0 | 321.45 | 69 | 323 | 120 | $1.01 \cdot 10^{10}$ |
| 7853 | 4618 | SB(rs)m | (R')SB(rs)m | Spitzer | 8.9 | 125.05 | 58 | 217 | 62 | $1.44 \cdot 10^9$ |
| 7876 | 4635 | SAB(s)d | SA(s)d | Spitzer | 14.5 | 61.25 | 53 | 344 | 98 | $1.44 \cdot 10^9$ |
| 7985 | 4713 | SAB(rs)d | SAB(rs)cd | Spitzer | 13.7 | 80.75 | 49 | 276 | 112 | $8.43 \cdot 10^8$ |
| 8403 | 5112 | SB(rs)cd | SB(s)cd | Spitzer | 19.1 | 119.45 | 57 | 121 | 120 | $1.54 \cdot 10^9$ |
| 8709 | 5297 | SAB(s)c | SAB$_x$(s)bc sp | Spitzer | 35.0 | 168.7 | 76 | 330 | 207 | $1.82 \cdot 10^{10}$ |
| 8852 | 5376 | SAB(r)b | | Spitzer | 30.6 | 62.7 | 52 | 63 | 186 | $1.47 \cdot 10^{10}$ |
| 8937 | 5430 | SB(s)b | (R$_1$')SB(s,nl)b | Spitzer | 49.0 | 65.65 | 32 | 185 | 275 | $1.73 \cdot 10^{10}$ |
| 9179 | 5585 | SAB(s)d | | Spitzer | 5.7 | 172.65 | 36 | 49 | 111 | $5.50 \cdot 10^9$ |
| 9358 | 5678 | SAB(rs)b | (R$_1$'L)SAB(rs)b pec | Spitzer | 29.1 | 99 | 54 | 182 | 221 | $2.39 \cdot 10^{10}$ |
| 9366 | 5668 | SA(rs)bc | S$\underline{A}$B(rs)c | Spitzer | 37.7 | 119.45 | 62 | 225 | 241 | $3.44 \cdot 10^{10}$ |
| 9465 | 5727 | SABdm | | Sloan | 26.4 | 67.15 | 65 | 127 | 97 | $7.27 \cdot 10^8$ |
| 9736 | 5874 | SAB(rs)c | | Sloan | 45.4 | 68.75 | 51 | 219 | 192 | $1.27 \cdot 10^{10}$ |
| 9753 | 5879 | SA(rs)bc | SAB(rs)bc | Spitzer | 12.4 | 125.05 | 69 | 3 | 138 | $4.34 \cdot 10^9$ |
| 9943 | 5970 | SB(r)c | SB(s)c | Spitzer | 28.0 | 86.5 | 54 | 266 | 185 | $1.16 \cdot 10^{10}$ |
| 9969 | 5985 | SAB(r)b | SAB(s)ab | Sloan | 36.0 | 164.85 | 61 | 16 | 311 | $4.73 \cdot 10^{10}$ |
| 10075 | 6015 | SA(s)cd | SAB$_a$(s)cd | Spitzer | 14.7 | 161.1 | 62 | 210 | 168 | $5.96 \cdot 10^9$ |
| 10359 | 6140 | SB(s)cd pec | SB(s)d | Spitzer | 16.0 | 189.3 | 44 | 284 | 143 | $7.14 \cdot 10^9$ |
| 10470 | 6217 | (R)SB(rs)bc | (R')SB(rs)b | Spitzer | 21.2 | 90.6 | 34 | 287 | 164 | $1.38 \cdot 10^{10}$ |
| 10546 | 6236 | SAB(s)cd | SB(s)dm | Spitzer | 20.4 | 86.5 | 42 | 182 | 106 | $4.17 \cdot 10^9$ |
| 10564 | 6248 | SBd | | Spitzer | 18.4 | 94.85 | 77 | 149 | 75 | $1.50 \cdot 10^{10}$ |
| 11012 | 6503 | SA(s)cd | SAB(s)bc | Spitzer | 5.3 | 212.4 | 72 | 299 | 117 | $9.95 \cdot 10^9$ |
| 11124 | | SB(s)cd | | DSS | 23.7 | 75.35 | 51 | 182 | 96 | $8.17 \cdot 10^9$ |
| 11283 | IC1291 | SB(s)dm | | Spitzer | 31.3 | 54.6 | 34 | 120 | 173 | $7.36 \cdot 10^9$ |
| 11407 | 6764 | SB(s)bc | | Spitzer | 35.8 | 68.75 | 64 | 65 | 158 | $2.31 \cdot 10^{10}$ |
| 11557 | | SAB(s)dm | | Sloan | 19.7 | 65.65 | 29 | 276 | 105 | $8.51 \cdot 10^8$ |
| 11861 | | SABdm | | Spitzer | 25.1 | 104 | 43 | 218 | 181 | $1.27 \cdot 10^{10}$ |
| 11872 | 7177 | SAB(r)b | | Spitzer | 18.1 | 92.7 | 47 | 86 | 183 | $8.81 \cdot 10^9$ |
| 12276 | 7440 | SB(r)a | | DSS | 77.8 | 42.4 | 33 | 322 | 94 | $3.35 \cdot 10^{10}$ |
| 12343 | 7479 | SB(s)c | | Spitzer | 26.9 | 122.2 | 52 | 203 | 221 | $3.44 \cdot 10^{10}$ |
| 12754 | 7741 | SB(s)cd | (R$_2$')SB(s)cd | Spitzer | 8.9 | 130.95 | 53 | 342 | 123 | $1.69 \cdot 10^9$ |

**Notes.** Column (1) identifies the galaxy using the UGC classification; the galaxies are also named according the conventional NGC and IC classification in Column (2). Column (3) & (4) gives the morphological type according to RC3 and Buta at al. (2015). In column (5) appears the survey from which the image is taken. Columns (6) & (7) give, respectively, the distance of the object and its radius for the 25 B-band mag·arcsec$^{-2}$ isophote according to NED database. Columns (8) and (9) show,

respectively, the values of the inclination angle and the position angle of the line of nodes of the galaxy. The asymptotic rotational velocity determined from the rotation curves is listed in column (10). Lastly column (11) gives the estimated values of the stellar mass.

2. 2. The Data Analysis

In the present study we have measured morphological and kinematical properties of the bars and the disk galaxies that host the bars. The morphological parameters of the disk galaxies are the inclination angle, the geometrical center of the galaxy and the position angle of the line of nodes; these parameters are needed for deprojecting the images. We also measured the maximum rotation velocity of the ionized hydrogen as the only kinematical parameter of the disk as a whole. Concerning the morphological properties of the bars, we determined the bar length and the position angle of the bar. The set of the measured bar parameters includes the bar strength, the bar corotation radius and the pattern speed of the bar.

2.2.1. The Disk Properties

Several parameters that characterize the galactic disk are determined, such as the galactic center, the inclination angle, the position angle of the major axis of the disk galaxy and the asymptotic circular velocity, which is defined as the value of the circular velocity that the rotation curve tends to when this curve is almost flat, which occurs for large galactocentric radii. The rotation curve is calculated using the ROTCUR task of the astronomical package GIPSY. This software fits a tilted ring model (Begeman 1987) to the velocity field, so it is also possible to determine the geometrical parameters of the galaxy, by allowing one single parameter to vary freely while the others are kept fixed; this is then repeated with another parameter, and so on. The values of these kinematical properties: inclination angle, position angle and asymptotic velocity, can be found in Table 1, columns 7, 8, and 9, respectively. The uncertainty of the position angle is taken to be 2º for all galaxies, and the uncertainty for the inclination angle is taken as 7º. For the asymptotic velocity, we assume a relative uncertainty of 10%. We also include in Table 1, column 10, the values of $r_{25}$, which are taken from the NASA Extragalactic Database. The stellar masses of the galaxies, which are given in column (11) of Table 1, are taken from the NASA Sloan-Atlas (NSA[6]), the masses of those galaxies which are not found in the NSA database are estimated following the linear fit of the expression between an arbitrary color and the V-band mass-to-light ratio given by Wilkins et al. (2013):

$$log\Gamma_V = log\left[\left(M/M_\odot\right)/\left(L_V/L_\odot\right)\right] = p_1 \cdot (m_V - m_B) + p_2 \qquad (1)$$

where the relative magnitude in V is taken from NED database, and the color B-V is obtained from the Hyperleda database[7]. The uncertainties of the stellar mass are not included in Table 1, but can be estimated to be ~ 60% of the mass of the galaxy (Mendel et al. 2014).

2.2.2. The Bar Length

---

[6] http://www.nsatlas.org/data
[7] http://leda.univ-lyon1.fr/

We described in section 1 some of the techniques which have been employed to measure the bar length in the past. The method which we have chosen here we consider well adapted to the type of data used, mainly 3.6 μm images from Spitzer, supplemented with R-band SDSS images where Spitzer data were not available for the galaxies we had analyzed kinematically, and with a few DSS images where neither of the former types of observations was available. In this study, we adopt the technique developed by Erwin & Sparke (2003), which was applied in Erwin (2004) and discussed in detail in Erwin (2005). The technique consists in performing an ellipse fitting on the image of the galaxy, using a script based on the ELLIPSE task of the IRAF astronomical software package, in order to generate the radial dependence of the ellipticity and the position angle of ellipses. Initial values of the center, inclination, position angle and ellipticity are estimated by eye inspection of the image. We then determine three different measurements that characterize the bar length, $\rho_\epsilon$, $r_{min}$ and $r_{10}$, which are defined as follows: $\rho_\epsilon$ is the radius where the peak of ellipticity has its maximum while the position angle of the fitted ellipses is almost constant (Wozniak & Pierce 1991; Wozniak 1995); $r_{min}$ is the radial position of the first minimum in ellipticity just outside of the ellipticity peak; and $r_{10}$ indicates the radius where the position angle of the ellipses differs by at least 10º with respect to the angle measured at $\rho_\epsilon$. By definition, the two latter magnitudes, $r_{min}$ and $r_{10}$, are larger than the first one, which is taken as a lower limit of the bar size, so we define the upper limit of the bar length, $\rho_{bar}$, as the minimum of these two radii, i.e. $\rho_{bar}$ = min($r_{min}$,$r_{10}$), thus giving a bracketed value of the bar length, between the minimum and maximum defined here.

The position angle of the bar is determined as the position angle at the radius equal to $r_\epsilon$, this angle is needed for deprojecting the two measurements of the bar length, which are calculated according to the following expression:

$$\rho^{deproj} = \rho_{proj} \cdot \cos\theta_{bar} \cdot (tan^2\theta_{bar} \cdot sec^2 i + 1)^{1/2} \qquad (2)$$

where $\rho_{proj}$ is the measured, projected bar length (i.e. $\rho_\epsilon$ and $\rho_{bar}$), $\theta_{bar}$ is the position angle of the bar with respect to the position angle of the disk galaxy, and $i$ is the inclination angle of the galaxy. This expression is also used to calculate the corresponding uncertainties. A discussion on the effects of working on deprojected images to determine the bar length can be found in Gadotti et al. (2007).

So, following this procedure we give an upper and a lower limit for the bar length, in other words, the end of the bar should be placed between $[\rho_\epsilon^{deproj}, \rho_{bar}^{deproj}]$. However, in order to assign a single value to the bar length we define $r_{bar}$ as the mean value of the deprojected values of $\rho_\epsilon$ and $\rho_{bar}$ and the associated error is taken to be the half of the difference between these values. With an image of the galaxy, we have confirmed that the two limits to the bar length, which we calculate, bracket the bar size estimated by visual inspection, and we confirm that $r_{bar}$ is a reliable measurement of the radius of the bar. In Table 2 (columns 2 and 3) the deprojected values of $\rho_\epsilon$ and $\rho_{bar}$ in kpc are given.

In order to characterize the effect of using images of different surveys taken in such different passbands (3.6μm for Spitzer, R broadband for SDSS and 0.85μm bands for DSS2), we selected a small subset of 6 galaxies for which images from these three surveys are available, and we measured their bar lengths. In figure 1 we plot $r_{bar}$ values as measured with Spitzer images compared with those with SDSS images, and in the right panel we compare $r_{bar}$ from Spitzer and DSS2 images. In the two plots the line marks the 1:1 proportion. We can see that the Sloan images give values of the bar length that match well, within error bars, with those calculated with infrared images, but the bar lengths calculated from DSS2 images are only just compatible with the Sloan bar lengths, showing a higher dispersion with respect to the 1:1 line for the longer bars than for the shorter ones. As the only three galaxies analyzed with DSS images do not have large bars, we can take the DSS2 bar lengths as reliable.

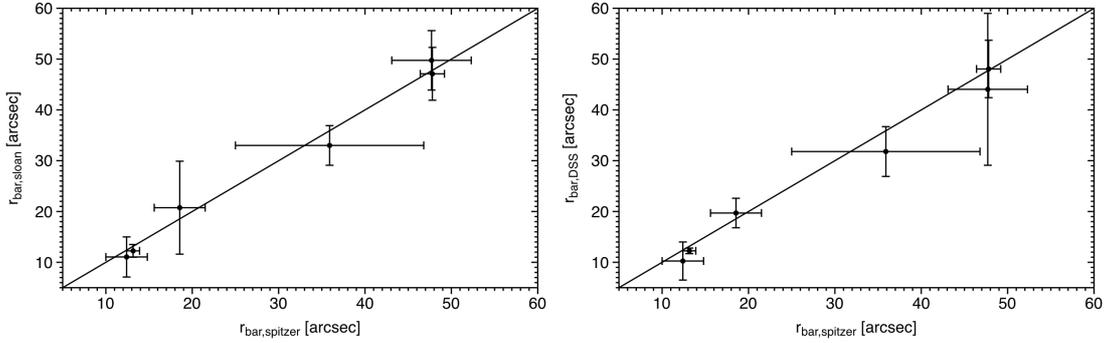

**Figure 1.** Comparison of the bar length measured from images of different surveys. **Left panel**. Results using infrared images are compared with those from SDSS in r band. **Right panel**. The same as left panel but using DSS images. The diagonal line plots the 1:1 relationship.

2.2.3. The Bar Strength

In the introduction we give an outline description of some of the methods which have been used to define and measure the bar strength. Here we measure the strength of the bar by performing a Fourier decomposition of the galaxy image. In the updated version of the kinemetry code (Krajnovic et al. 2006), the surface brightness image is written as a combination of a finite number of harmonic terms:

$$\Sigma(r,\varphi) = A_0(r) + \sum_{m=1}^{N} A_m(r) \cdot \sin(m\varphi) + B_m(r) \cdot \cos(m\varphi) \qquad (3)$$

where $\varphi$ is the azimuthal angle, and $r$ is the length of the semimajor axis of the elliptical ring in which the code performs the harmonic fitting. With this we calculate the harmonic coefficients $A_m$ and $B_m$ as a function of the radius. Although the main contribution to the bar strength comes from the amplitude of the harmonic term $m = 2$, Ohta et al. (1990) showed that the contribution of the even terms $m=4, 6$ is not negligible, so we include these higher order terms in the calculation of the bar strength, which is defined as the integration over a fixed radial range of the Fourier amplitude for the harmonics $m=2,4,6$ relative to the $m=0$ mode:

$$S_b = \frac{\sum_{m=2,4,6} \int_{r_1}^{r_2} \sqrt{A_m^2 + B_m^2}\, dr}{\int_{r_1}^{r_2} A_0\, dr} \qquad (4)$$

in which $r_1$ and $r_2$ are arbitrary limits that characterize the bar region. In our calculations we take the integration limit $r_1$ to be the half of the deprojected lower limit of the bar length $\rho_\epsilon$, and $r_2$ equal to the deprojected upper limit of the bar length $\rho_{bar}$, in order to restrict the region that is dominated by the bar, excluding, therefore, any contribution to the non-axisymmetric part from the spiral arms. To test our procedure we have measured the bar strength taking only the amplitude of the harmonic m=2, and we find a strong linear correlation between these values and the bar strength calculated using the expression (4). Kinemetry also gives the uncertainty associated with each harmonic term, so propagating these uncertainties using the expression (4), we can estimate the uncertainty of the bar strength. Doing this, we obtain approximately the same uncertainty of ~0.04 for all galaxies.

In order to quantify how reliable are the measurements of the bar strength when using an image in the R-band optical waveband from SDSS survey and in near-infrared from DSS2 survey ($\lambda_{eff}$ = 0.85 μm) with respect to the values calculated from an infrared image, we perform the calculations of $S_b$ of a subset of six random galaxies for which we have images in the three surveys. Results show that the bar strengths calculated from Sloan images do reproduce the values obtained from the Spitzer images, see left panel of Figure 2; the data show a small dispersion with respect to the 1:1 solid line. On the other hand, the $S_b$ values from DSS2 images are really far from the infrared values, as shown in the right panel of Figure 2. When we examine the numerator of expression (4) we find that that the values for DSS2 are lower (nearly the half) than those measured with the corresponding Spitzer images. Furthermore, while in the central regions the radial dependence of the term A0 (the denominator in expression (4)) is similar for the Spitzer and the DSS2 images, but further out A0_spitzer decays as a function of the radius, while A0_DSS is almost uniform. To cover this case, a linear fit to the data was made (plotted as a dashed line), and this is used to correct for the strength of the bar measured with DSS images:

$$S_b^{Spitzer} = 0.009 + 3.125 \cdot S_b^{DSS} \qquad (5)$$

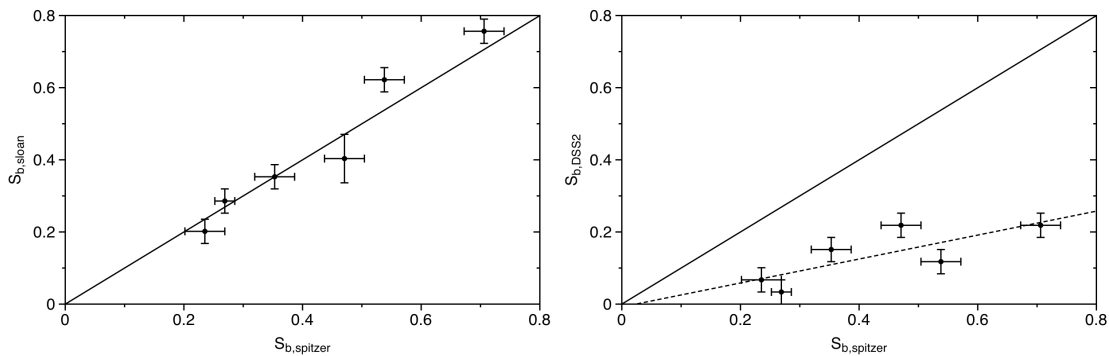

**Figure 2**. Comparison of bar strength values calculated from three types of images. Solid line indicate the 1:1 relationship. **Left panel**. Sloan values are compared with those from Spitzer. **Right panel**. DSS values are compared with the Spitzer values, the dashed line plots the linear fit to the data.

The calculated values of the bar strength and their associated uncertainties are listed in column 6 of Table 2.

2.2.4. The Corotation Radius of the Bar and the Pattern Speed

In the introduction we give an outline review of the methods which have been used in the past to determine the corotation radius for a bar, either via measurements of the pattern speed, or by measuring the corotation radius itself. In this article we determine the bar corotation radius (and hence its pattern speed) with the Font-Beckman method (Font et al. 2011, 2014a), which uses phase reversals of the streaming motions. In essence, from the line of sight velocity field in Hα we extract the rotation curve, which is used to construct a 2D model of the circular velocity, which is then subtracted from the original velocity field in order to produce the residual velocity map. On this map we identify those pixels where the non-circular velocity switches in sign, the phase-reversals, for which we can calculate their deprojected radii. With this we can plot the radial distribution of the phase-reversals and each peak in this distribution can be associated with a resonance. In particular, the strongest peak found in the bar region is assigned to the bar corotation; this has also been successfully applied to double barred galaxies in Font et al. (2014b). The corresponding angular velocities, the pattern speeds of the bars, are then determined using the frequency curves derived from the rotation curve. The values of these derived parameters describing the bar, along with their uncertainties are given in Table 2, columns 6-9.

2.2.5. The Scaled Properties

We present the measured properties scaled by a corresponding parameter characterizing the disk galaxy. So, in what follows, the bar length is given relative to the radius of the galaxy for the $25^{th}$ magnitude isophote, $r_{25}$. Following this scheme, we define the parameter $\Gamma$ as the bar pattern speed, $\Omega_{bar}$, divided by the angular velocity of the disk, $\Omega_{disk}$, which is calculated as the asymptotic rotation velocity divided by the radius $r_{25}$ (in kpc). The values of this scaled parameter are in column 8 of Table 2.

Measuring the corotation radius of the bar relative to the bar length, we can determine the rotational parameter, $\mathcal{R}$, which is conventionally used to distinguish between fast rotator and slow rotator bars, depending on whether the value of $\mathcal{R}$ is lower or higher than 1.4, respectively (Debattista & Sellwood 2000). It is important to emphasize that the designation "fast" ("slow") rotator bar does not mean that the bar is rotating with a high (low) angular speed; this classification is based only on the value of the rotation parameter, $\mathcal{R}$. Contopoulos (1980) studied how a galaxy responds to the barred perturbations of density, and he concluded that the bar corotation should be found close to the end of the bar, and in principle the value of the rotational parameter should be quite close to, but larger than 1. This general conclusion has since been supported in diverse publications, both theoretical and observational, (Athanassoula 1992; Valenzuela & Klypin 2003; Rautiainen et al. 2008; Buta & Zhang 2009; Font et al., 2014a; Aguerri et al. 2015). As described in a former section, we give the two limits for the bar length, consequently we calculate the two limits for the $\mathcal{R}$ parameter, the values of this parameter appearing in Table 2, column 7, are the calculated as the mean of the two limits.

The formation and evolution of the bar in disk galaxies is studied by means of purely N-body or hydrodynamical simulations in which, under different initial assumptions in terms of secular evolution or interaction effects, the time variation of the bar length, the bar strength, the pattern speed and the rotational parameter are calculated. Most numerical simulations take into account only the stars, but with the increasing power of the computers more recent simulations include gas particles (Friedli & Benz 1993, 1995; Patsis & Athanassoula 2000; Bournaud & Combes 2002; Berentzen et al. 2007; Villa-Vargas et al. 2010) and the effect of the dark matter halo (Debattista & Sellwood 2000; Athanassoula & Misiriotis 2002; Valenzuela & Klypin 2003; Villa-Vargas et al. 2009, 2010), showing that these play a significant role in the bar evolution process.

**Table 2.**

Bar parameters of the galaxies in the sample.

| Name UGC | $\rho_\epsilon$ (arcsec) | $\rho_{bar}$ (arcsec) | $r_{CR}$ (arcsec) | $\Omega_P^{bar}$ (km·s$^{-1}$·kpc$^{-1}$) | $S_{bar}$ | $\mathcal{R}$ | $\Gamma$ |
|---|---|---|---|---|---|---|---|
| (1) | (2) | (3) | (4) | (5) | (6) | (7) | (8) |
| 508 | 46.8±2.0 | 49.7±2.0 | 51.8±2.3 | $32.6^{+1.3}_{-1.2}$ | 0.37 | 1.07±0.07 | 1.61±0.20 |
| 763 | 40.3±2.2 | 46.7±2.1 | 51.9±1.9 | 26.2±0.6 | 0.30 | 1.20±0.04 | 1.90±0.11 |
| 1256 | 25.2±2.2 | 57.4±2.3 | 58.7±3.0 | 26.5±1.0 | 0.17 | 1.68±0.09 | 2.37±0.15 |
| 1317 | 15.5±2.1 | 17.2±2.1 | 26.3±5.1 | $34.7^{+6.2}_{-4.6}$ | 0.45 | 1.61±0.31 | 4.64±0.73 |
| 1437 | 11.0±2.0 | 11.2±2.0 | 16.4±5.6 | $40.8^{+18.4}_{-10.5}$ | 0.10 | 1.47±0.50 | 4.57±1.63 |
| 1736 | 30.2±3.5 | 56.0±5.1 | 56.1±2.1 | 33.5±0.9 | 0.15 | 1.43±0.05 | 2.08±0.37 |
| 1913 | 95.4±2.0 | 117.3±2.0 | 134.5±4.6 | 16.2±0.8 | 0.22 | 1.28±0.04 | 2.18±0.20 |
| 2080 | 23.4±2.0 | 29.8±2.1 | 32.1±1.3 | $46.1^{+1.3}_{-1.2}$ | 0.35 | 1.22±0.05 | 3.21±0.52 |
| 2855 | 42.8±2.5 | 44.0±2.5 | 70.4±6.3 | $33.5^{+2.9}_{-2.7}$ | 0.30 | 1.62±0.15 | 1.63±0.14 |
| 3013 | 72.3±6.3 | 105.1±8.2 | 100.1±2.7 | 10.4±0.3 | 1.21 | 1.17±0.03 | 1.21±0.07 |
| 3463 | 28.5±2.7 | 32.0±2.8 | 45.5±7.5 | $19.1^{+3.1}_{-2.4}$ | 0.25 | 1.51±0.25 | 1.93±0.28 |
| 3685 | 23.2±2.0 | 29.5±2.0 | 27.7±1.7 | 22.3±0.3 | 0.40 | 1.07±0.07 | 2.77±0.23 |
| 3709 | 16.6±2.7 | 21.8±2.9 | 24.6±1.7 | $27.0^{+1.6}_{-1.5}$ | 0.54 | 1.31±0.09 | 1.59±0.10 |
| 3740 | 17.2±3.0 | 19.4±3.1 | 33.3±1.6 | 19.9±0.5 | 0.30 | 1.82±0.09 | 1.60±0.19 |
| 3809 | 41.7±4.3 | 53.2±5.2 | 62.0±4.9 | $24.6^{+1.8}_{-1.6}$ | 0.11 | 1.33±0.10 | 3.23±0.23 |
| 4165 | 32.5±4.1 | 41.2±4.8 | 48.9±2.0 | $27.3^{+1.0}_{-0.9}$ | 0.34 | 1.35±0.06 | 1.57±0.19 |
| 4273 | 32.2±3.2 | 44.6±4.0 | 47.4±3.2 | $22.4^{+1.3}_{-1.1}$ | 0.25 | 1.27±0.09 | 1.35±0.08 |
| 4325 | 58.6±2.0 | 62.1±2.0 | 67.2±3.9 | 18.5±0.6 | 0.22 | 1.11±0.06 | 1.20±0.10 |
| 4422 | 38.6±2.8 | 43.6±2.9 | 44.3±1.5 | $29.2^{+1.0}_{-0.9}$ | 0.55 | 1.08±0.04 | 2.21±0.30 |
| 4555 | 9.3±2.0 | 10.4±2.0 | 13.0±2.6 | $41.1^{+6.3}_{-4.9}$ | 0.15 | 1.32±0.26 | 2.97±0.47 |
| 4936 | 38.2±2.0 | 43.0±2.0 | 70.8±2.1 | $24.1^{+0.6}_{-0.5}$ | 0.22 | 1.75±0.05 | 2.46±0.14 |
| 5228 | 17.7±2.0 | 31.1±2.0 | 30.4±1.3 | $32.3^{+1.1}_{-1.0}$ | 0.45 | 1.35±0.06 | 2.28±0.11 |
| 5303 | 24.7±2.0 | 27.8±2.0 | 35.2±6.2 | $56.9^{+10.4}_{-7.7}$ | 0.11 | 1.35±0.24 | 2.69±0.43 |
| 5319 | 10.4±2.0 | 13.1±2.0 | 16.2±1.2 | $51.3^{+2.5}_{-2.3}$ | 0.30 | 1.40±0.10 | 2.46±0.34 |
| 5510 | 14.7±2.0 | 17.6±2.0 | 30.8±1.6 | $50.9^{+2.0}_{-1.9}$ | 0.49 | 1.92±0.10 | 2.49±0.34 |
| 5532 | 6.9±2.3 | 11.1±2.3 | 16.3±3.3 | $115.9^{+26.6}_{-18.8}$ | 0.13 | 1.92±0.39 | 6.77±1.34 |
| 5786 | 19.3±4.9 | 23.1±4.9 | 20.6±2.1 | $37.1^{+9.3}_{-6.2}$ | 0.49 | 0.98±0.10 | 2.96±0.68 |
| 5840 | 29.8±2.1 | 35.2±2.1 | 60.9±5.4 | $84.6^{+6.7}_{-5.8}$ | 0.12 | 1.89±0.18 | 2.39±0.19 |
| 5842 | 15.5±2.1 | 22.0±2.2 | 23.4±1.5 | 39.0±0.8 | 0.17 | 1.29±0.08 | 2.26±0.19 |
| 5982 | 14.0±2.0 | 16.3±2.0 | 19.5±2.0 | $73.3^{+31.4}_{-16.1}$ | 0.17 | 1.29±0.13 | 4.44±1.44 |
| 6118 | 33.3±2.0 | 43.9±2.0 | 43.7±1.7 | 58.9±1.9 | 0.57 | 1.15±0.05 | 1.90±0.11 |
| 6537 | 44.4±2.1 | 53.9±2.1 | 59.8±2.5 | 35.4±0.8 | 0.24 | 1.23±0.05 | 2.43±0.12 |
| 6778 | 17.8±2.0 | 20.5±2.0 | 34.9±4.2 | $63.4^{+6.2}_{-5.1}$ | 0.12 | 1.83±0.22 | 2.86±0.27 |
| 7021 | 25.3±5.1 | 29.3±5.6 | 30.0±3.2 | $48.2^{+6.7}_{-5.6}$ | 0.86 | 1.11±0.12 | 2.89±0.40 |
| 7154 | 38.5±3.8 | 44.7±4.1 | 63.4±2.7 | 19.8±0.5 | 0.71 | 1.53±0.07 | 1.89±0.08 |
| 7323 | 56.4±3.4 | 61.6±3.6 | 92.4±7.4 | 18.6±0.9 | 0.47 | 1.57±0.13 | 1.31±0.13 |
| 7420 | 31.6±2.3 | 36.1±2.3 | 36.1±3.1 | $49.6^{+3.2}_{-2.9}$ | 0.44 | 1.07±0.09 | 5.26±0.34 |
| 7766 | 17.9±3.7 | 27.9±5.1 | 37.3±1.9 | $39.4^{+1.5}_{-1.4}$ | 0.17 | 1.71±0.09 | 6.65±0.35 |
| 7853 | 23.5±3.2 | 43.7±3.3 | 41.7±1.1 | 19.4±0.5 | 0.27 | 1.36±0.04 | 1.69±0.48 |
| 7876 | 16.4±3.8 | 21.1±4.2 | 28.7±1.9 | $36.4^{+1.3}_{-1.2}$ | 0.20 | 1.55±0.10 | 1.60±0.13 |
| 7985 | 19.5±3.2 | 21.4±3.3 | 38.3±5.3 | $38.3^{+4.4}_{-3.7}$ | 0.29 | 1.88±0.26 | 1.83±0.22 |
| 8403 | 10.6±2.0 | 25.4±2.1 | 30.8±2.0 | 20.2±0.4 | 0.32 | 2.05±0.13 | 1.86±0.09 |
| 8709 | 33.0±2.0 | 44.4±2.0 | 50.0±1.4 | 24.1±0.5 | 0.54 | 1.32±0.04 | 3.33±0.10 |
| 8852 | 17.4±2.0 | 22.0±2.0 | 26.8±1.3 | $44.5^{+1.8}_{-1.7}$ | 0.12 | 1.38±0.07 | 2.23±0.11 |
| 8937 | 20.2±2.2 | 36.6±2.3 | 33.3±3.0 | $35.9^{+4.0}_{-3.4}$ | 1.06 | 1.28±0.12 | 2.04±0.44 |
| 9179 | 61.6±2.1 | 78.1±2.1 | 74.3±1.5 | 44.7±1.1 | 0.30 | 1.08±0.02 | 1.92±0.32 |
| 9358 | 8.4±2.0 | 17.3±2.0 | 14.7±1.9 | $92.4^{+10.3}_{-8.4}$ | 0.37 | 1.30±0.17 | 5.84±0.62 |
| 9366 | 25.4±2.5 | 28.1±2.6 | 36.9±5.6 | $34.1^{+5.6}_{-4.3}$ | 0.25 | 1.38±0.21 | 3.09±0.45 |
| 9465 | 10.4±1.6 | 13.1±1.6 | 21.5±2.0 | 26.7±1.2 | 0.20 | 1.85±0.17 | 2.37±0.15 |
| 9736 | 8.3±1.4 | 9.0±1.4 | 12.1±2.2 | $46.6^{+7.1}_{-5.2}$ | 0.34 | 1.40±0.26 | 3.67±0.51 |
| 9753 | 15.9±2.0 | 38.7±2.0 | 30.6±1.9 | $74.6^{+5.0}_{-4.4}$ | 0.22 | 1.36±0.08 | 4.06±0.29 |
| 9943  | 16.2±2.0 | 22.5±2.0  | 34.4±4.2  | $39.5^{+4.9}_{-4.0}$   | 0.27 | 1.83±0.22 | 2.51±0.29 |
| 9969  | 15.9±1.4 | 19.0±1.4  | 27.7±2.8  | $50.2^{+2.5}_{-2.4}$   | 0.42 | 1.60±0.16 | 4.64±0.24 |
| 10075 | 10.3±2.0 | 14.3±2.0  | 16.1±2.3  | $71.6^{+3.5}_{-3.2}$   | 0.17 | 1.34±0.19 | 4.89±0.26 |
| 10359 | 73.3±2.0 | 90.1±2.0  | 112.8±1.4 | 14.8±0.2               | 0.50 | 1.40±0.02 | 1.52±0.16 |
| 10470 | 35.2±2.5 | 54.8±2.9  | 57.8±3.4  | $24.9^{+1.4}_{-1.3}$   | 0.92 | 1.35±0.08 | 1.41±0.18 |
| 10546 | 13.7±2.0 | 15.4±2.0  | 28.1±1.5  | 36.3±1.5               | 0.10 | 1.93±0.11 | 2.93±0.33 |
| 10564 | 32.8±2.6 | 42.2±2.8  | 45.6±1.5  | 12.4±0.3               | 0.67 | 1.23±0.04 | 1.40±0.08 |
| 11012 | 16.9±4.0 | 17.8±6.3  | 21.8±1.8  | 107.7±2.0              | 0.35 | 1.26±0.10 | 5.02±0.21 |
| 11124 | 22.7±3.1 | 31.3±3.7  | 48.3±2.7  | 14.4±0.3               | 0.30 | 1.84±0.10 | 1.30±0.10 |
| 11283 | 21.9±2.0 | 26.1±2.0  | 38.4±2.6  | $21.9^{+2.3}_{-1.9}$   | 0.72 | 1.62±0.11 | 1.05±0.24 |
| 11407 | 49.6±2.0 | 56.3±2.0  | 64.9±2.3  | 12.9±0.3               | 0.40 | 1.23±0.04 | 1.07±0.10 |
| 11557 | 21.7±1.6 | 26.0±1.6  | 32.4±2.6  | 18.4±0.4               | 0.22 | 1.37±0.11 | 1.10±0.38 |
| 11861 | 28.5±3.7 | 33.1±4.1  | 37.8±3.7  | $24.8^{+1.1}_{-1.0}$   | 0.34 | 1.23±0.12 | 1.73±0.20 |
| 11872 | 17.2±3.3 | 19.3±3.5  | 19.9±2.8  | $94.5^{+11.9}_{-9.6}$  | 0.77 | 1.09±0.15 | 4.20±0.50 |
| 12276 | 14.2±2.3 | 18.0±2.4  | 21.8±1.3  | 11.9±0.6               | 0.17 | 1.37±0.08 | 2.02±0.41 |
| 12343 | 63.5±5.0 | 108.4±7.8 | 91.1±4.3  | 18.4±1.0               | 0.94 | 1.14±0.05 | 1.33±0.08 |
| 12754 | 47.3±7.0 | 77.0±10.0 | 69.6±7.3  | $36.6^{+2.4}_{-2.1}$   | 0.50 | 1.19±0.12 | 1.68±0.13 |

**Notes.** Columns (2) and (3) give the upper and lower limit, respectively, for the deprojected bar length of the galaxies named in column (1), as described in the text. The bar corotation radius in arcsec and the pattern speed of the bar calculated with the Font-Beckman method are given in columns (4) and (5), respectively. The calculated values of the bar strength appear in column (6) with an uncertainty of 0.04 for all galaxies. The rotational parameter, defined as the bar corotation length scaled by the bar size is given in column (7). Column (8) shows the angular velocity of the bar in units of the angular velocity of the disk.

3. RESULTS AND DISCUSSION

In this section we analyze the parameters which characterize the bar. We start with the distributions of each magnitude calculated, and then we study how these parameters are related to the morphological type and the total stellar mass of the hosting galaxy.

3.1. Statistics of the Bar Parameters

With the data listed in Table 2, we can construct the histograms of each measured parameter. We plot in Figure 3, the distribution of the rotational parameter (left panel) and the relative bar pattern speed (right panel). The histogram of the relative bar corotation shows clearly that the rotational parameter is somewhat larger than unity, within uncertainties, as predicted theoretically by Contopoulos (1980), who studied how a bar responds to the barred perturbations of density, deriving the result that the bar corotation resonance must occur beyond, but close to the bar end. The mean value of this parameter and its standard deviation for our sample is $\bar{\mathcal{R}} = 1.41 \pm 0.26$, which is compatible within uncertainties with $\bar{\mathcal{R}} = 1.2 \pm 0.2$ given by Athanassoula (1992). We also distinguish clearly four separated peaks in the histogram, the central positions of the peaks are found at: $\mathcal{R}_1 = 1.09$, $\mathcal{R}_2 = 1.32$, $\mathcal{R}_3 = 1.59$ and $\mathcal{R}_4 = 1.85$, so the first two peaks are due to bars which are traditionally termed fast rotators, as the limiting value to classify bars as fast/slow rotators is 1.4 (Debattista & Sellwood 2000). Adopting this classification we find that two thirds of our galaxies host a fast rotator bar.

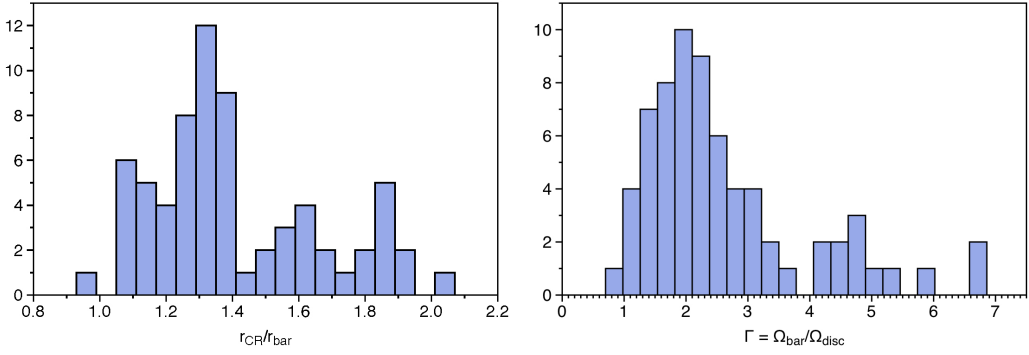

**Figure 3.** Histogram of the rotational parameter (left panel), and the relative bar pattern speed (right panel).

It is widely accepted, from both observations and modeling, that the rotational parameter increases from early to late type galaxies (Aguerri et al. 1998; Rautiainen et al. 2008; Font et al., 2011, 2014a) and this could induce a misinterpretation of the histogram of this parameter (left panel of figure 3) in which each peak could be erroneously associated with a particular morphological type of galaxy if only the mean value of this parameter is taken into account. With the values of $\mathcal{R}$ appearing in column 7 of Table 2, we can see that in each peak we have a contribution of galaxies of many different morphological types. We can see in Figure 3, right panel, that the four peaks in the dimensionless parameter $\mathcal{R}$ have no equivalent in the distribution of the relative bar pattern speed, which shows two different populations: the bars that rotate over four times faster than the disk (a minority), and the majority of the galaxies whose bar has an angular rate lower than four times the angular speed of the disk, with a mode value of twice the angular speed. In any case, we find that all bars rotate faster than their disks.

3.2. Variation of Bar Parameters with the Morphological Type of the Galaxy

In this section we study the dependence of the bar properties, which are listed in Table 2, on the morphological type of the host galaxy. In Figure 4 we plot this variation for the relative bar length (panel a), the bar strength (panel b), the rotational parameter (panel c) and the relative bar pattern speed (panel d). In all plots, the red squares mark the mean value of the parameter for the specific morphological type and the standard errors of the mean are indicated as error bars.

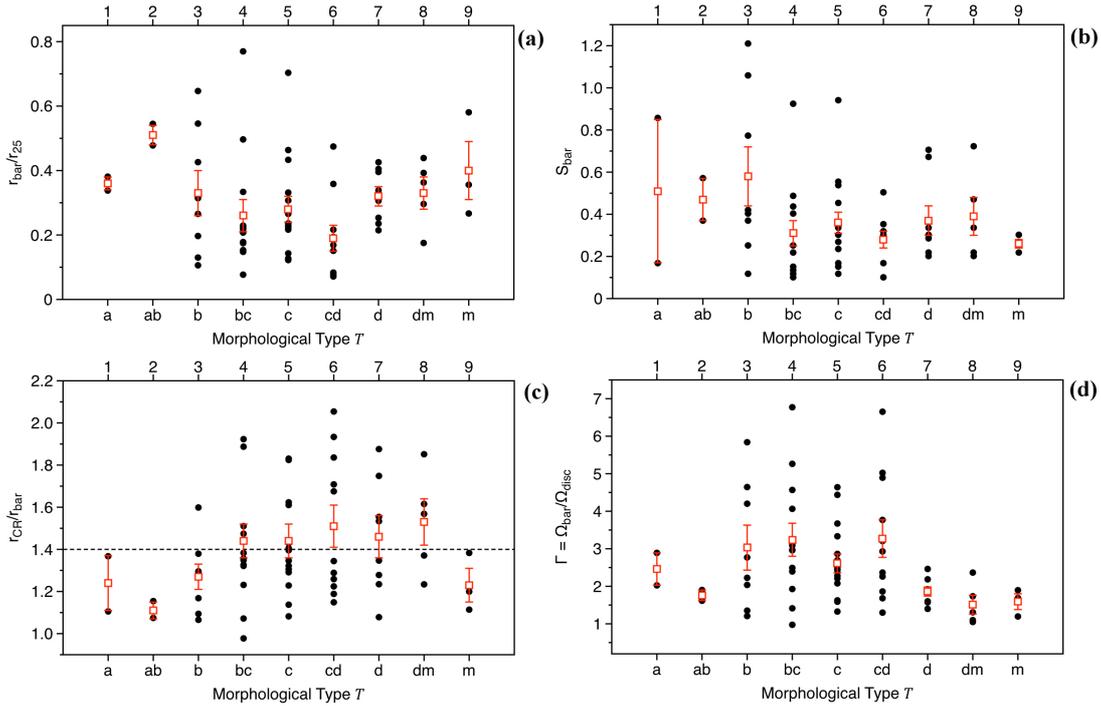

**Figure 4.** Distribution of the bar parameters as function of the morphological type of the host galaxy: The relative bar length in panel (a); the bar strength in panel (b); the bar corotation radius relative to the bar length in panel (c), the horizontal dashed line marks the separation between fast/slow rotators as conventionally defined; and in panel (d) the scaled pattern speed of the bar. The red boxes mark the mean value of the parameter for a specific morphological type galaxy, and the errors bars show the standard error of the mean.

In the top left panel of Figure 4 we see that the earlier galaxies of type $T$=1,2 host the largest bars, then the relative bar size of the bar drops from galaxies of ab morphological type down to cd-type of galaxies ($T = 6$), for which the shorter bars are found, and this is followed by a clear increase through the later type of galaxies. When we compare our bar length distribution as a function of the galaxy morphological type with the same distribution found in other studies, we find that in general they are in agreement, although our sample of 68 spiral galaxies is statistically poor especially for the earlier types; all these distributions show a local minimum of the relative bar length for galaxies of morphological type $T = 6$. This was also found by Martin (1995) who studied 136 spiral galaxies, Laurikainen et al. (2007), who analyzed a sample of 216 galaxies, and Díaz-García et al. (2016) with a very large sample of more than 600 galaxies belonging to the survey S4G.

The bar strength does not show large variations through galaxies of morphological type $T \geq 4$ (see top right panel of Figure 4). In Laurikainen et al. (2007), the distribution of bar strength as a function of galaxy morphological type is plotted for a sample of 216 galaxies. When the bar strength is calculated from the torque maps, they find a steady growth along the Hubble sequence. The same method was employed by Díaz-García et al. (2015) with a large sample of ~600 galaxies, finding a similar result. Similarly, Seidel et al. (2015) using torques to estimate the bar strength, also find this growth of the bar strength from early to intermediate type of galaxies, with a sample of 16 galaxies.

The variation of the rotational parameter is displayed in panel (c) of Figure 4, where the horizontal dashed line marks the border between fast and slow rotators. The figure shows a significant rise of this parameter from early type galaxies to intermediate type galaxies and then the parameter remains almost constant until a large drop for the SBm galaxies, although the numbers here are small. The mean value of the ratio between bar corotation radius and the bar length and its standard deviation for early-type galaxies (SBa, SBab) is found to be $\bar{\mathcal{R}}_{early} = 1.18 \pm 0.12$, for intermediate type galaxies (SBb, SBbc) is $\bar{\mathcal{R}}_{inter} = 1.37 \pm 0.27$, and for late-type galaxies $\bar{\mathcal{R}}_{late} = 1.45 \pm 0.25$. These results indicate that the corotation to the bar radius ratio shows a slight tendency to grow from morphological earlier to late type galaxies, as also pointed out in Aguerri et al. (1998), but these values are also consistent with the conclusion that this parameter shows no dependence on the morphological type of the galaxy, when the standard deviations are taken into account. The mean values of the rotational parameter for the three morphological types of galaxies determined in the present study are in agreement with those found in Font et al. (2014a) with a sample of smaller size, see Table 3. However, in that study the sizes of the bars were taken from the literature, which means that they were calculated by different authors using different methods. Modeling infrared images, Rautiainen et al. (2008) derived values of the corotation over the bar radius shown in Table 3, which are in agreement, within the uncertainties, with the values obtained from our sample. In disagreement with the values shown in Table 3, Aguerri et al. (2015), applying the Tremaine-Weinberg method to determine the pattern speed of 15 galaxies from the CALIFA survey plus 17 from the previous literature, found that this parameter shows no significant variation along the Hubble sequence.

**Table 3.**
The rotational parameter for different morphological type of galaxies

| Reference | $\bar{\mathcal{R}}_{early}$ | $\bar{\mathcal{R}}_{inter}$ | $\bar{\mathcal{R}}_{late}$ | N |
|---|---|---|---|---|
| Rautiainen et al. 2008 | 1.15±0.25 | 1.44±0.29 | 1.82±0.63 | 38 |
| Font et al. 2014a | 1.15±0.28 | 1.30±0.30 | 1.35±0.28 | 32 |
| This study | 1.18±0.12 | 1.37±0.27 | 1.45±0.25 | 68 |

**Notes.** The mean values, and their standard deviations, of the corotation to the bar length ratio for earlier, intermediate and later type of galaxies are given in the three central columns. In the first column appears the reference from which these values are taken, and in the last column the size of the sample is listed.

In the last plot of Figure 4 (panel d), we show the variation of the scaled bar pattern speed with the morphological type of galaxy, we find that the bar of an earlier type galaxy ($T = 1,2$), rotates slowly, then the parameter jumps and remains uniform for galaxies of Hubble type between 3 and 6, and finally the bar angular rate drops to low values for galaxies with $T \geq 7$. Numerical simulations of bar evolution predict the slowdown of the bar (Winberg, 1985; Athanassoula, 2003), so this braking of the bar from intermediate ($3 \leq T \leq 6$) to later type of galaxies ($T \geq 7$) we show in the distribution could be interpreted as a sign of galaxy evolution only if we assume that late type galaxies are more evolved than intermediate type galaxies, which is potentially interesting but beyond the scope of the present article.

3.3. Dependence on the Mass of the Galaxy

We now show how the properties of the bar are affected by the stellar mass of the host galaxy. To do so, we plot the bar length (in kpc), the bar strength, the bar corotation scaled to the bar size and the relative bar angular rate as a function of the stellar mass of the galaxy (in units of solar masses) in Figure 5, from panel (a) to (d), respectively.

In Kormendy (1979) it was shown that the bar size is well correlated with the galaxy mass (absolute magnitude); this dependence is shown in panel (a) of Figure 5. Although the total stellar mass is determined with rather large uncertainties, we show in that figure that the largest bars can be found only in the most massive galaxies, while the shortest bars are hosted in the intermediate and low mass galaxies, this tendency is also found in Díaz-García et al. 2016.

In the plot of the stellar mass of the galaxy versus the bar strength (panel (b) of Figure 5), we can see that all data points are uniformly distributed in the upper half confined by a lower limiting diagonal in this parametric plane; consequently the lower half is a forbidden region. Interpreting the extent of the permitted region, we conclude that the strongest bars are present only in massive galaxies (although the most massive galaxies can also have weaker bars), and that the bars within the less massive galaxies can only be the weaker ones. Somewhat similar behavior is found in panel (d) where the galaxy masses are plotted against the bar angular speed. The figure shows that the bars that rotate fastest are only in those galaxies with intermediate mass (~ $10^{10}$ solar masses), in other words, we do not detect any galaxy in our sample having a very low or a very high stellar mass that hosts a bar that spins with an angular speed greater than ~ 50 km s$^{-1}$kpc$^{-1}$, these fastest bars are only present in galaxies with an intermediate mass.

The plot of the stellar mass of the galaxies against the rotational parameter is shown in panel (c) of Figure 5. We do not find any particular trend between these two parameters, so if we assume the conventional classification of bars as fast or slow rotators depending on whether $\mathcal{R}$ is below or above 1.4, respectively (this critical value is marked in the plot as a vertical dashed line), we find that the two type of bars are present in galaxies regardless of their stellar mass.

All data plotted in the four panels of figure 5 are colored according to the morphological type of the galaxy; Early type galaxies (in blue) include galaxies of type a & ab, intermediate type (in red) include b & bc type galaxies, and late type (in green) all the remaining galaxies. So, comparing the intermediate and late type galaxies, as the number of early type galaxies is of relatively lower significance, we can infer from panels a & d of figure 5 that late type galaxies are less massive than the intermediate type, and the bars in late type galaxies are short, rotating with low angular rates compared with the bars hosted in intermediate type galaxies. We cannot compare these two types of galaxies in terms of bar strength and rotational parameter as the population of these galaxies appear blended in the diagrams of panels b & c of figure 5.

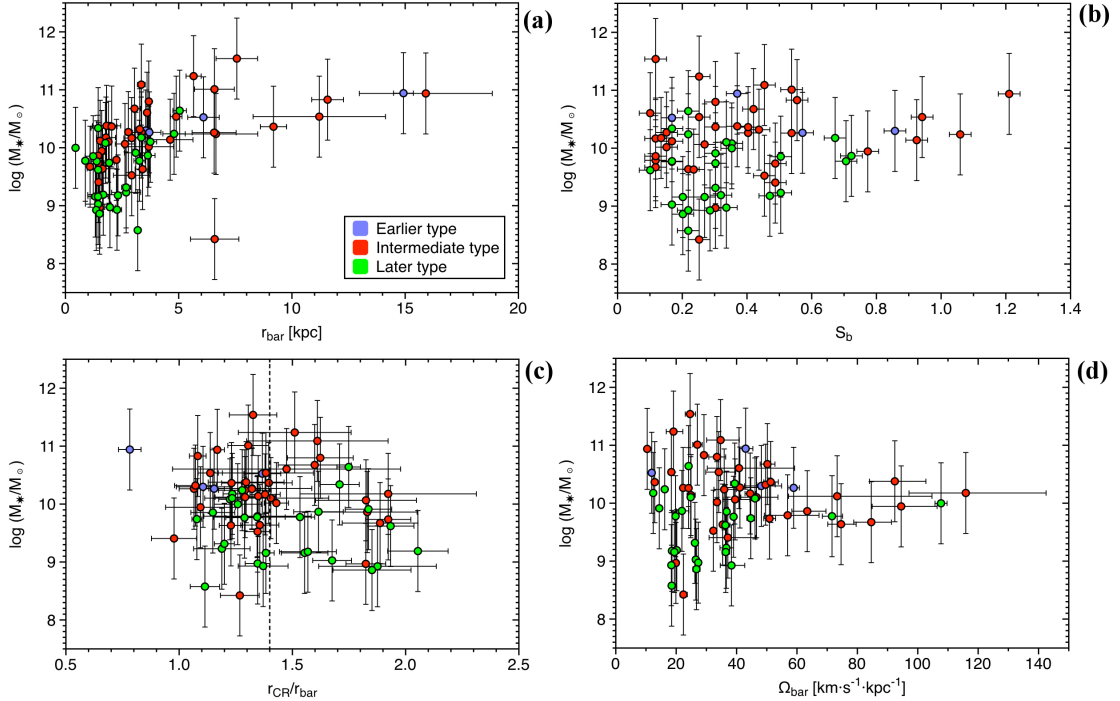

**Figure 5.** Plots of key bar parameters plotted against the stellar mass of the galaxies, in units of solar masses. Data in all panels are color-coded according to the morphological type of the galaxy (see legend in panel a). **Panel (a).** The bar length, in kpc. **Panel (b).** The bar strength. **Panel (c).** The ratio between the corotation radius and the bar size. The vertical dotted line indicates the critical value used in the literature to distinguish between fast and slow rotating bars. **Panel (d).** The pattern speed of the bar obtained with the Font-Beckman method.

## 4. RELATIONSHIP BETWEEN BAR PARAMETERS

In this section we investigate how the different parameters characterizing the bar are related to each other. The results of numerical simulations of bar formation and evolution are used to help interpret the plots we produce in this section. From Martínez-Valpuesta et al. (2017) we have picked out two different simulations; in both cases the galaxy develops a bar and 43% of the baryonic mass within a radius of 7 kpc is in the disk. In the first simulation (I1_d_500 in Martínez-Valpuesta et al. 2017), which is named here as model 1, there is a fly-by interaction with a second galaxy that occurs at $t = 1.45$ Gyr. In model 2 (I1_d_2000 in Martínez-Valpuesta et al. 2017), the evolution of the bar is governed only by internal processes. A third model (model 3) is also considered, this model is similar to model 1 but with a different fraction of baryonic mass in the disk, which now is 92%. The main properties of the three models are summarized in Table 4. These simulations are taken as qualitative guideline models rather than being used for specific predictions.

**Table 4.**
Properties of the numerical simulations

| Model | Interaction | Baryonic mass |
|---|---|---|
| Model 1 | Fly-by | 43% |
| Model 2 | No interaction | 43% |
| Model 3 | Fly-by | 92% |

**Notes.** All models listed in column 1 are correspond to numerical simulation by Martínez-Valpuesta et al. (2017). The third column gives the fraction of baryonic mass in the disk within a radius of 7 kpc.

4.1. The Bar Pattern Speed versus the Bar Length

In Figure 6 we combine panel (a) and (d) of Figure 5, by plotting the bar angular rate against the bar length. We define three different groups of galaxies depending on their total stellar mass, and use this classification to color the data in the plot. We can see that each group of galaxies occupies a defined region in this ($r_{bar}$-$\Omega_{bar}$) parametric space, so we infer from this figure that: (*i*) The longest bars ($r_{bar} \gtrsim 10$ kpc) can rotate only with low angular speed ($\Omega_{bar} \lesssim 40$ km·s$^{-1}$·kpc$^{-1}$) and can be found only in the most massive galaxies (blue points in the figure). (*ii*) Those short bars ($r_{bar} \lesssim 3$ kpc), which rotate slowly ($\Omega_{bar} \lesssim 40$ km·s$^{-1}$·kpc$^{-1}$), are present only in the less massive galaxies (data in red of Figure 6). (*iii*) The bars that rotate with highest angular speed, are necessarily short ($r_{bar} \lesssim 3$ kpc), and can be hosted only in galaxies of intermediate mass (green points in the top-left region of Figure 6). In general, figure 6 shows that those bars shorter than ~ 3 kpc can be present only in galaxies of small or medium mass, depending on whether the pattern speed is lower or higher than ~ 40 km·s$^{-1}$·kpc$^{-1}$, respectively. Moreover we find, for the smallest bars, a linear correlation between the bar pattern speed and the total stellar mass of the galaxy, as shown in Figure 7, in which red dots correspond to those bars with radial length less than 3 kpc and the blue dots correspond to the remaining bars with $r_{bar} > 3$ kpc (uncertainty bars are omitted for clarity). The solid line in this figure shows the linear fit performed when only the smallest bars are considered (red dots), which follows the expression:

$$\log\left(\frac{M_*}{M_\odot}\right) = 7.21 + 1.50 \cdot \log\left(\Omega_{bar}[km \cdot s^{-1} \cdot kpc^{-1}]\right)$$

Figure 7 also demonstrates that we would find no overall relationship between these parameters if all bars were taken together.

It is known from many N-body simulations of barred galaxy evolution that as the galaxy evolves the bar tends to grow and also to suffer a deceleration in its angular rotation rate (Athanassoula 2003, 2013; Berentzen et al. 2007; Villa-Vargas et al. 2009, 2010; Martínez-Valpuesta et al. 2006, 2017). The galaxy as a whole also experiences a growth in mass, which has been monitored by the star formation and the Active Galactic Nuclei activity (see section 15.3 of Mo et al. 2010), and in particular the stellar mass of the galaxy in a dark matter halo grows while the galaxy is evolving (van de Voort 2016, and references therein). According to this scenario, an evolved barred galaxy is characterized by having a large stellar mass, and by harboring a slowly rotating, and long bar (galaxies located in the bottom-right region of Figure 6) compared with the initial values of these parameters prior to its evolution (top-left region in Figure 6). The simulations of bar evolution show how these two stages are connected, as can be seen in Figure 8 (circles for model 1, and squares for model 2) where we plot the trajectory that a bar follows in the ($r_{bar}$,$\Omega_{bar}$) parametric plane after the buckling of the bar, obtained from two simulations of the bar evolution by Martínez-Valpuesta et al. (2017). The evolution time, for 3 Gyr, proceeds from the red points (early stages) to the orange points (most evolved stages), the points are connected by a solid line to bring out the trajectory. We see that the two trajectories displayed in Figure 8 describe the slowdown and the growth of the bar, but for model 2 (squares), the path to lower pattern speeds and larger bar lengths is a very wide 'S' shape, which is not seen in the case of the fly-by interaction (model 1) which is better described by a 'decay-type' trajectory. So, the same galaxy at different evolutionary stages can populate different parts of the diagram.

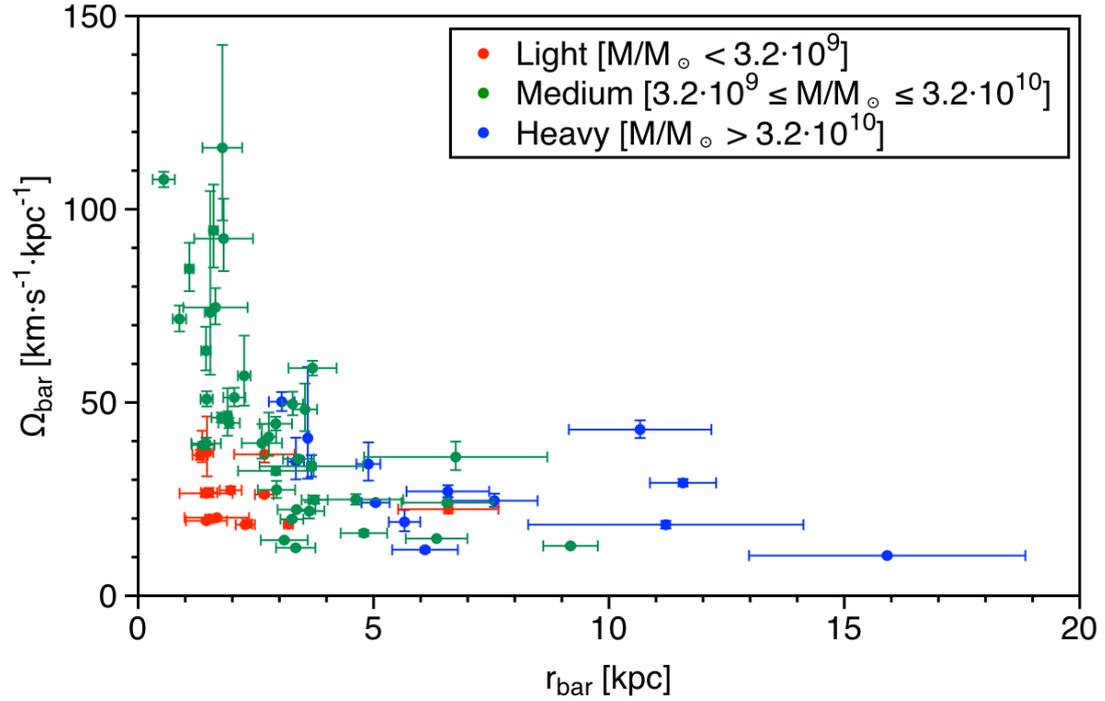

**Figure 6.** Plot of the bar pattern speed, in km·s$^{-1}$·kpc$^{-1}$, versus the bar length, in kpc. The data are colored according to the total stellar mass of the galaxy.

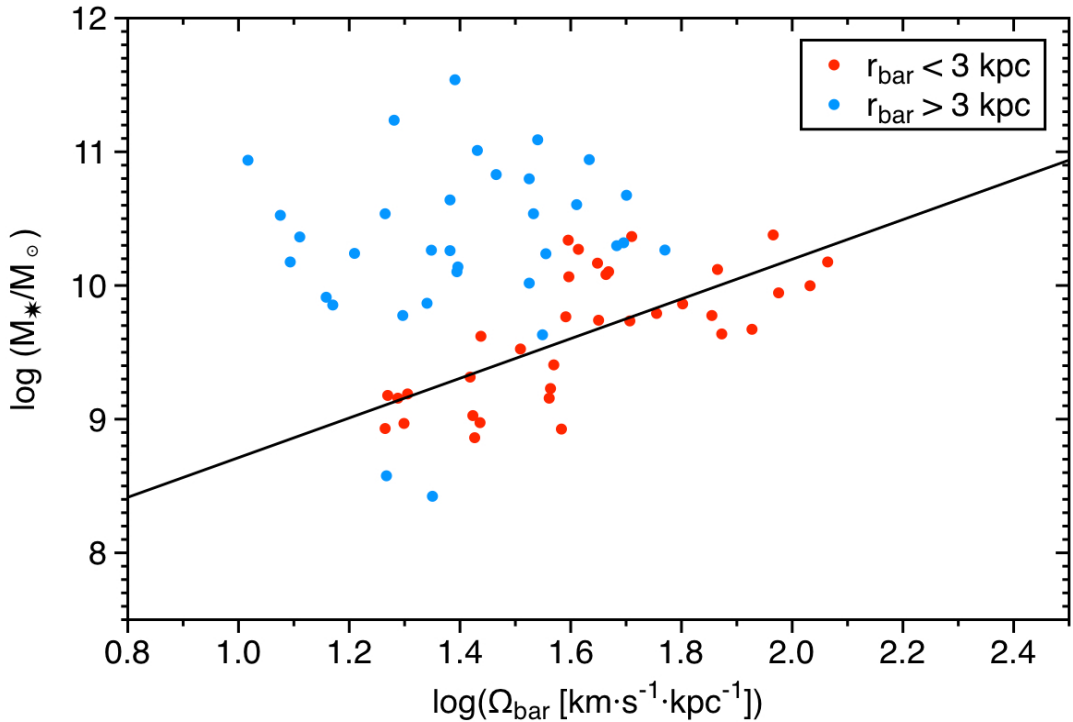

**Figure 7.** The total stellar mass of the galaxy (in units of solar mass) versus the bar pattern speed (in units of km·s$^{-1}$·kpc$^{-1}$), on logarithmic scale. The red/blue dots represent the bar with a size lower/higher than 3 kpc. The solid line shows the linear fit performed for only the shortest bars.

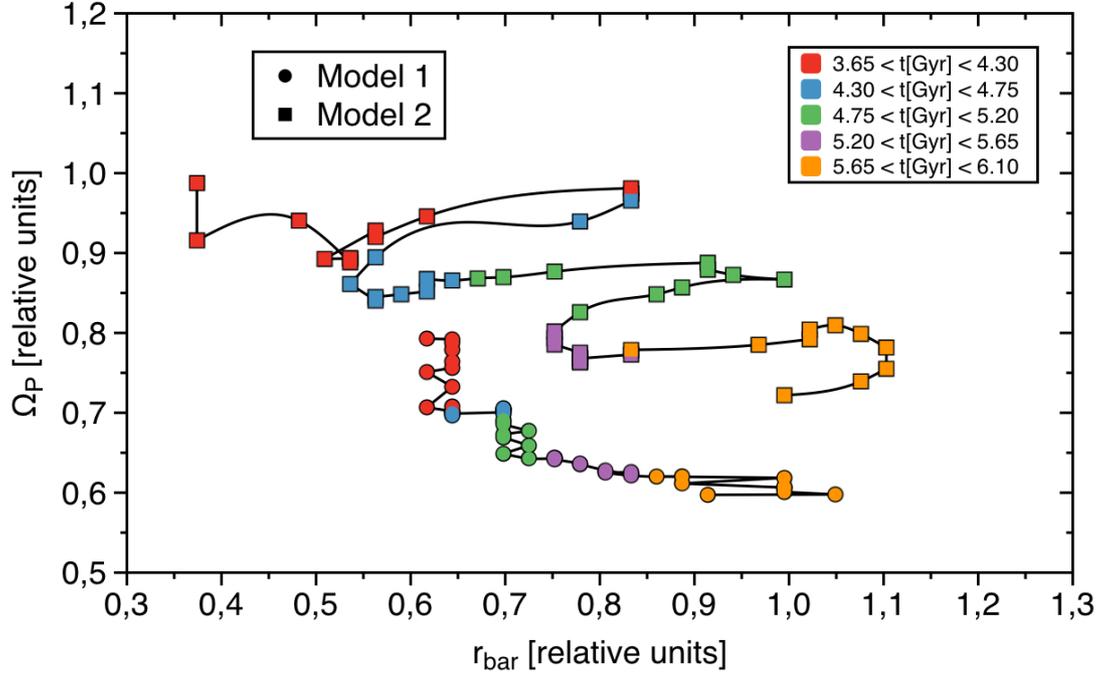

**Figure 8.** Evolution of the bar length (in arbitrary units) and the bar pattern speed (in arbitrary units) found in the N-body simulations by Martínez-Valpuesta et al. (2017) for a barred galaxy which interacts with a fly-by galaxy (model 1, in circles) and for model 2 (in squares). The plot displays the trajectory in this parametric plane in which time runs implicitly from red points to orange ones (as indicated in the table of colors), the points are connected with solid lines, showing that the bar evolves by increasing its size while it is slowing down.

It would not be valid to identify directly the trajectories in the ($r_{bar}$,$\Omega_{bar}$) parametric plane obtained from numerical simulations of a single galaxy with our data plotted in Figure 6 but we can use these trajectories to infer some general trends of bar evolution from this figure. Although the decay trajectory in the ($r_{bar}$,$\Omega_{bar}$) plane for an interacting galaxy (circles in Figure 8), seems to reproduce in shape the distribution of our data in Figure 6 reasonably better than the 'S'-shaped trajectory for the non-interacting galaxy, we cannot conclude that the fly-by interaction is the preferred scenario for the bar formation, as numerically the bar of model 2 is shorter and shows larger values of the pattern speed in the early stages, compared with the bar of model 1. However, both mechanisms of bar production are broadly compatible with our observational measurements. Reading the data in figure 6 in terms of the trajectories for the two models plotted in figure 8, we see that the data on the left side of figure 6 (galaxies with a short bar), can be associated with the early stages of barred galaxies, while the data in the bottom-right region of figure 6 are consistent with the most evolved galaxies.

4.2. The Scaled Bar Angular Frequency versus the Bar Strength

The bar strength is another magnitude studied in the numerical simulations by tracing its evolution. In general, all simulations predict that the bar strength grows while the bar grows in size and reduces its angular velocity. With our measurements we can produce the plots of the bar strength versus the bar pattern speed relative to the disk velocity, and versus the relative bar size.

In Figure 9, we plot the relative bar pattern speed against the bar strength, coloring the data according the relative bar length. First we can see from Figure 9 that the scaled bar size is an appropriate parameter to distribute the data, as it is possible to distinguish the regions in the ($\Gamma$-$S_b$) parametric plane occupied by the short bars, the medium bars and the large bars. The largest bars are located in the bottom region of the plot, where the bars rotate more slowly, while the shortest bars are at the top, where the relative angular rate is higher. We also can notice that all points are confined to half of this parametric plane, as they are distributed only over the bottom triangle defined by the inverted diagonal, or to be explicit we do not find any galaxy in our sample that harbors a strong bar rotating with a higher angular speed than the disk. We see from this figure that the strongest bars are those that are spinning with the lowest pattern speed (bottom-right region), while the bars that rotate faster can only have lower values of the bar strength (top-left region).

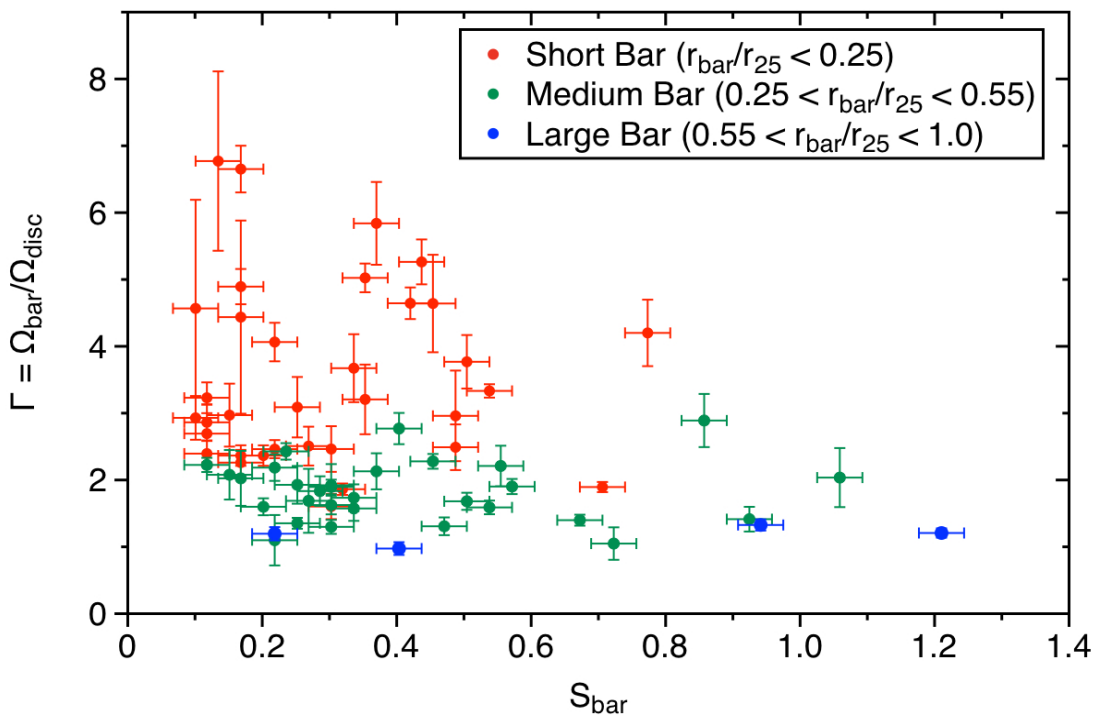

**Figure 9.** Plot of the bar pattern speed scaled by the disk angular velocity, versus the bar strength. The data are color coded for relative bar length.

It is not straightforward to see trends of bar slowdown and bar strength growth in Figure 9. Again, numerical simulations can provide a useful aid to the interpretation of our results in terms of bar evolution. In Figure 10 we plot the time evolution of these two magnitudes, as trajectories in the parametric plane, for three different simulations; in addition to the two models already described (model 1 & 2), we also include here a third model (model 3, diamonds in Figure 10, left panel) in which there is no interaction, as in model 2, but with a different distribution of matter in the disk; while in model 2 a 43% of baryonic matter is in the disk within a radius of 7 kpc (thus the remaining 57% in this part of the disk is the contribution of the dark matter), in model 3 the fraction of matter within a radius of 7 kpc of the disk is 92%, so there is almost no dark matter in that region of the disk. We denote the bar strength obtained in the numerical simulations as $A_2$ as it is calculated as the integrated Fourier amplitude of only the *m=2* mode normalized by the *m=0* mode. The values of the bar strength, $A_2$ from the simulations can be qualitatively compared with those we calculate using the amplitude of even higher modes, as there is a strong linear correlation between the bar strength calculated using expression (4) and calculated only considering the contribution of the *m*=2 mode. In Figure 10, the evolution time flows from the green points to the grey ones, in the same way as in Figure 9, so Figure 10 displays how the bar slows down and its bar strength increases while the bar evolves. Additionally, the bar growth can be clearly seen in the right panel of figure 10, where the same data is color coded according to the relative size of the bar, showing that, regardless of the model, the bar is shorter in the early stages of the evolution (red points), and becomes larger after 3 Gyr of evolution (blue points). In the right panel of Figure 10, the bar of model 2 (squares) is left to develop for an extra time of 0.9 Gyr with respect to the same bar plotted in the left panel, in order to extend the evolution of these parameters, showing that they keep the same correlation found in the left panel. We can recognize the same pattern in Figure 9, reading the points of that figure in diagonal (trajectories in Figure 10), so given one diagonal in Figure 9, the shortest bars and least evolved ones are those rotating with the highest pattern speed and having the lowest values of the bar strength, while the more evolved bars have larger bar strengths and bar lengths, and rotate more slowly; this is also reproduced in the two simulations of the right panel of Figure 10.

It is clear from the left panel of Figure 10 that the variation of some parameters of the numerical simulations (such as the mass distribution in the disk, or the contribution of dark matter, or the fact of considering an interaction with a second galaxy) can extend the coverage of the simulations in the diagram, so our numerical results plotted in figure 9 could be used to constrain the initial values of these parameters for further simulations of bar evolution. The results of the numerical simulations seem to indicate that those galaxies found in the short diagonal (close to the origin of the parametric plane) should contain a very low fraction of dark matter in the disk as model 3, with only 8% of dark matter within the region out to 7kpc radius in the disk, reproduce the range of values of bar pattern speed and bar strength that we have found in our measurements. On the other hand, those galaxies on the large diagonal (farthest from the origin of the parametric space) seem to contain a significant contribution of dark matter in that region of the disk, which can be estimated to be larger than ~ 60% of the total mass, as model 2 simulates the evolution of a bar with values of these parameters close to that large diagonal.

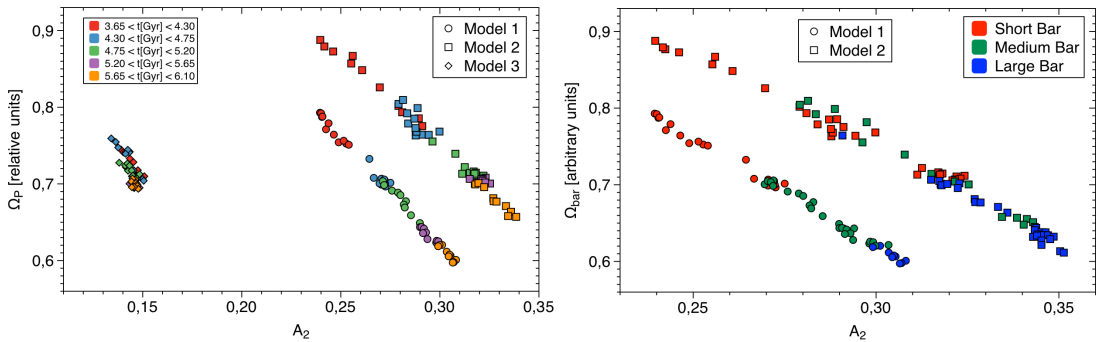

**Figure 10. Left panel.** Trajectory of the bar evolution in the parametric plane of bar pattern speed and bar strength according to numerical simulations of Martínez-Valpuesta et al. (2017) for the model 1, in which the bar formation is triggered by an interaction with a secondary galaxy (circles), for model 2, which forms the bar by internal processes (squares), and for model 3, which is similar to model 2 but with a different mass distribution of the disk. The evolution time runs from the red points to the orange ones, as indicated in the color scale. **Right panel.** Same as left panel, but only for model 1 and 2. In this plot the points are colored according to the relative bar size.

4.3. The Bar Length versus the Bar Strength

In Figure 11, we plot the relative bar length versus the bar strength for the barred galaxies of our sample. In the plot, the data is coded in colors according to the morphological type of the galaxy, i.e. the earlier, intermediate and later type, galaxies of type {a,ab,b}, {bc,c,cd} and {d,dm,m}, respectively. First we can see that although there is not a clear correlation between these magnitudes, the strongest bars do tend to be the longest ones, while the shortest bars tend to have lower values of the bar strength. The plot also shows that the morphological type does not play an important role in determining these parameters, as the colored points are well mixed.

We have used numerical simulations in order to help give a possible interpretation to our results, so in Figure 12 we plot the evolution of the bar size and the bar strength, corresponding to model 1 and 2, in the left and right panel, respectively. The simulations indicate that the evolution of the two parameters is more or less correlated depending on the model considered, so for the bar of model 1 (left panel of Figure 12) we see a strong correlation between the bar strength and the bar length, while the data of model 2 in the parametric plane ($r_{bar}$,$S_{bar}$) show a higher dispersion (right panel of Figure 12). In any case, in both models, when the bar is evolving, the bar strength increases while the bar is growing in size. This tendency between the bar strength and the bar length is also found in Figure 11, where the largest bars tend to be the strongest ones, and should be the most evolved bars according to the simulations (trajectories in Figure 12). Numerically, model 2 develops a larger and a stronger bar compared with model 1, but in the early stages of the formed bar it is shorter and rotating more slowly in model 1 than in the non-interacting model 2. Thus, our measurements do not, at this stage, favor one scenario of the bar formation between the two scenarios taken into account here (i.e. with and without interaction corresponding to model 1 and 2, respectively). The numerical simulations also offer an interpretation of our results plotted in Figure 11 in terms of more/less evolved bars; the former are strong and large bars, while the latter are shorter and weaker bars.

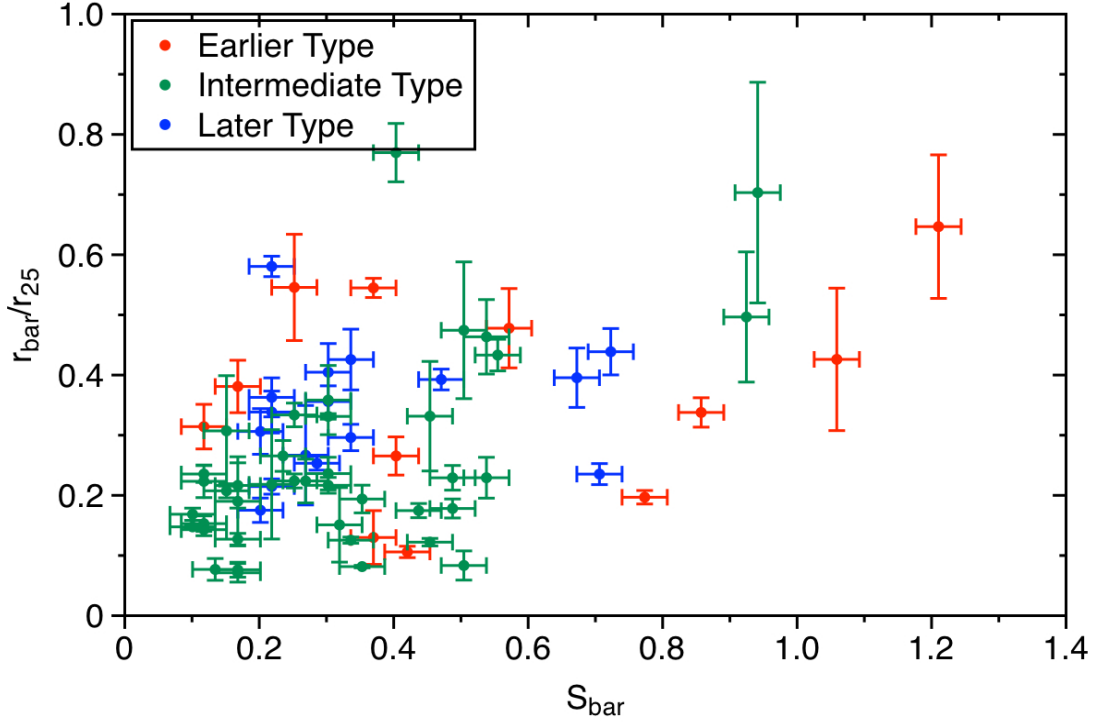

**Figure 11**. Plot of the relative bar size versus the bar strength obtained for the galaxies of our sample, which are classified depending on its morphological type as early (red), intermediate (green) and late type (blue).

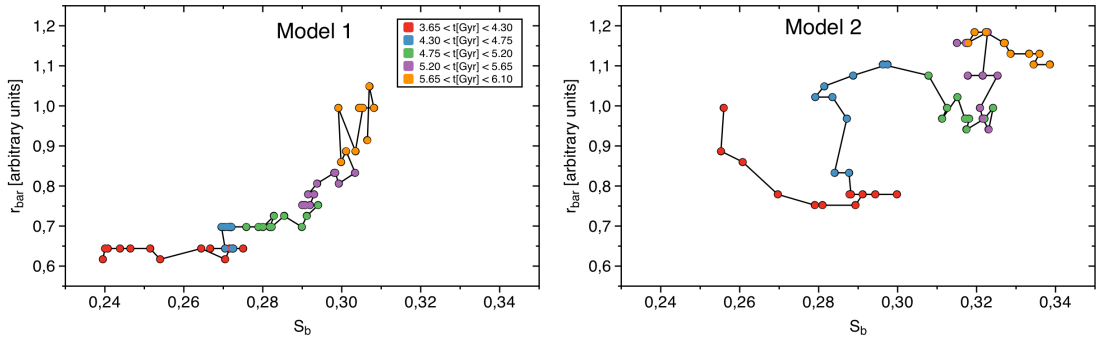

**Figure 12. Left panel.** Plot of the evolution of the bar length and the bar strength obtained from numerical simulations of Martínez-Valpuesta et al. 2017 for the model 1, in which the bar is formed by an interaction. The evolution time flows from the red points to the orange ones as shown in the color scale. **Right panel.** Same plot as left panel but for the model 2 in which the bar is formed only by gravitational instabilities. The time evolution scale is the same than in the left panel.

5. WHY DO "FAST ROTATOR" BARS ROTATE MORE SLOWLY THAN "SLOW ROTATOR" BARS?

A widely accepted criterion to distinguish between fast rotator bars and slow rotator bars is whether the bar corotation to the bar length ratio is lower or higher than 1.4, respectively. As we have measurements of the bar pattern speed and the rotational parameter for all galaxies of our sample, we plot these two parameters in Figure 13 in which we mark the limiting value $\mathcal{R} = 1.4$ with a vertical dashed line. The plot shows that in the region of the "fast rotator" bars ($\mathcal{R} \leq 1.4$), we find the same range of values for the scaled pattern speed as that found in the region of "slow rotator" bars. In other words, the fastest bars are found in the fast rotator region, which is expected, but also in the slow rotator region. And the same happens with those bars that have the lowest values of the pattern speed; they are found in both regions (fast rotator and slow rotator region). As we don't find any correlation between the two parameters of Figure 13, it seems more reasonable to use the relative angular speed of the bar as the suitable parameter to distinguish between fast and slow bars. So we can classify the bars as fast/slow rotators depending on whether the scaled pattern speed is over/below to 2.0, i.e. whether the bar is rotating more or less quickly than twice the speed of the disk.

The relative bar size does give a well organized plot for the data in Figure 13, in which points are colored according to this parameter, showing that the largest bars (in blue) are those that rotate slower with values of the bar pattern speed relative to the disk angular speed below 2.0, and also having the lowest values of the rotational parameter, below 1.4 which would make them fast rotators according to the usual classification. The smallest bars (in red) are those having a higher angular rate ($\Gamma > 2.0$, and are thus fast rotator bars with our definition) and covering the full range of values of the ratio between the corotation radius and the bar length. We find that all the short bars of our sample rotate faster than the largest bars. As we can distinguish three different regions in the $\Gamma - \mathcal{R}$ parametric plane depending on the bar size, we calculate the mean value of these two parameters along with the standard deviation for each group of galaxies, obtaining the following values: $\bar{\mathcal{R}}_{large} = 1.16 \pm 0.05, \bar{\Gamma}_{large} = 1.18 \pm 0.15$, for the galaxies with a large bar, $\bar{\mathcal{R}}_{int} = 1.35 \pm 0.20, \bar{\Gamma}_{int} = 1.84 \pm 0.44$, for galaxies with an intermediate bar size, and $\bar{\mathcal{R}}_{short} = 1.51 \pm 0.28, \bar{\Gamma}_{short} = 3.55 \pm 1.38$ for galaxies having a short bar. These results show that the bar pattern speed and the rotational parameter decrease from the shortest bars to the largest ones, as DO their standard deviations, showing that the cloud of points for short bars is more scattered than that for intermediate bars, and that this cloud is more scattered than that for the largest bars. It is interesting to note that mean values of the ratio between the bar corotation radius and its length for large, intermediate and short bars, are compatible with those found for early, intermediate and late morphological type of galaxies, respectively (see Table 3).

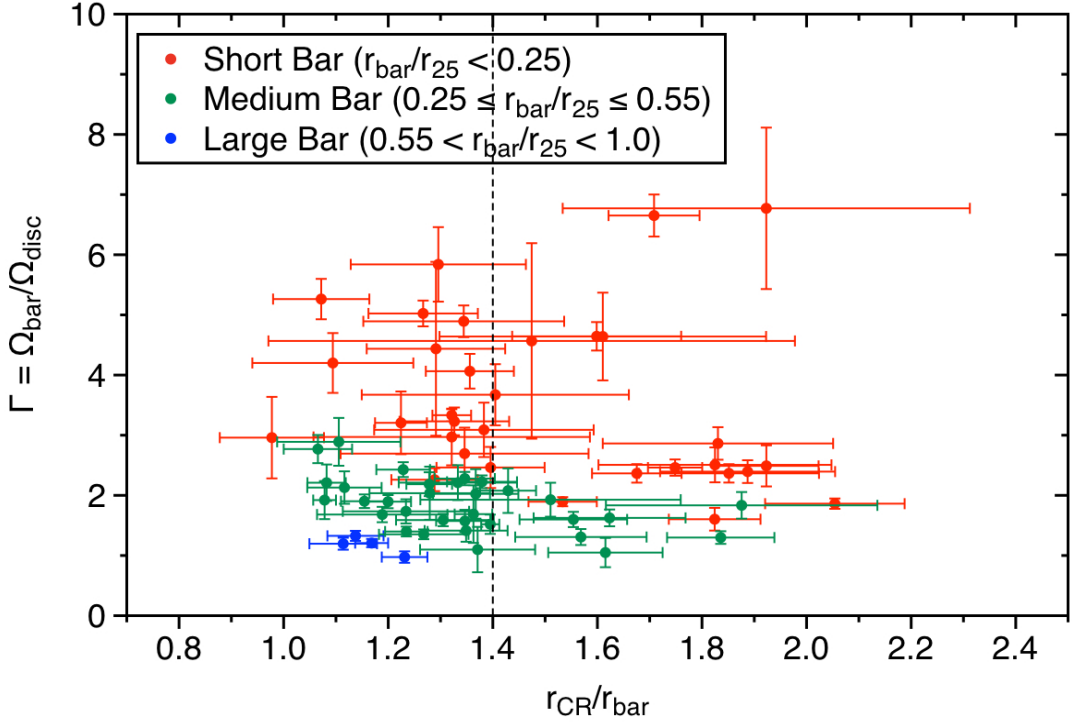

**Figure 13**. Plot of the bar pattern speed relative to the disk angular rate versus the ratio between the corotation radius and the bar length, the data is colored according to the relative bar size.

5.1. So, What do we really Mean when we Use the Terms Fast/Slow Rotator Bar?

The bar pattern speed and the bar corotation radius are linked by means of the angular velocity curve, which is calculated from the rotation curve by dividing the circular velocity by the galactocentric radius. In Figure 14, left panel we show a typical rotation curve of a disk galaxy, and in the right panel we plot the derived angular velocity curve in which the relationship between the pattern speed and the corotation radius is illustrated in red at two different radial positions of the bar resonance. This latter panel illustrates the conventional definition of a fast/slow rotator bar, which has to be understood as follows: given a bar with a bar length, $r_{bar}$, marked as solid vertical line in the right panel of Figure 14, we define the critical corotation position at 1.4 times $r_{bar}$, this is marked in the plot as a vertical dotted line, which defines two different radial regions; between $r_{bar}$ and $r_{critical}$ filled in green in the plot (fast rotator range), and beyond $r_{critical}$ in blue (slow rotator range). If the corotation of the bar occurs within the green region, marked as $r_{CR}^F$, then the corresponding pattern speed is $\Omega_{bar}^F$, but if the if the corotation radius falls in the blue region ($r_{CR}^S$ in the figure), then the corresponding pattern speed is $\Omega_{bar}^S$, and it is obvious from the right panel of Figure 14 that $\Omega_{bar}^S < \Omega_{bar}^F$, therefore, the bar rotates faster when the corotation radius is in the green region compared to the same bar for which the resonance is found in the blue region. So, when corotation occurs between $r_{bar}$ and 1.4 times $r_{bar}$, that bar is conventionally called a fast rotator bar (as first defined in Debattista & Sellwood 2000), while it is classified as a slow rotator bar if the resonance is located beyond 1.4·$r_{bar}$. However this terminology of fast/slow rotator bars turns out to be confusing when we compare different bars for which we also know their pattern speeds, for example UGC1913 and UGC9969. The former galaxy has a bar with a pattern speed of 16.2 km·s$^{-1}$·kpc$^{-1}$ and the ratio between corotation and bar length is 1.28, so it's a fast rotator bar. The bar of UGC9969 is rotating with an angular speed of 50.2 km·s$^{-1}$·kpc$^{-1}$, and $\mathcal{R}$=1.60, so is classified as a slow rotator. So, we have a slow rotator bar (UGC9969), which is spinning faster than a fast rotator bar (UGC1913).

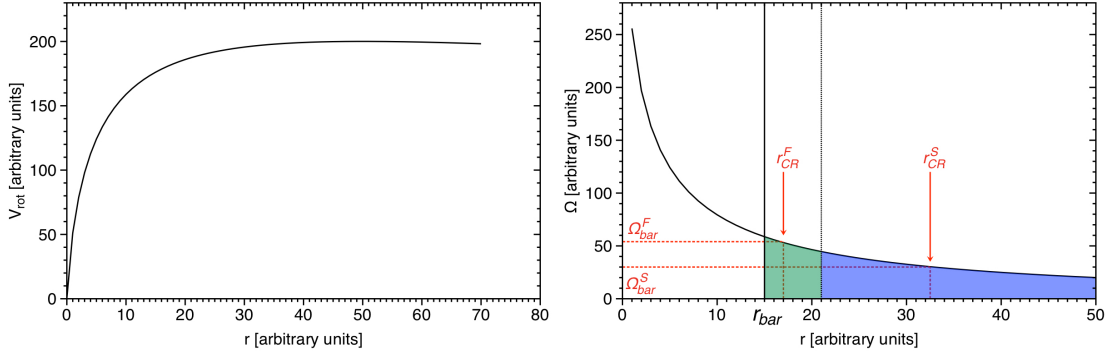

**Figure 14. Left panel.** Schematic plot of the typical rotational curve of a disk galaxy. **Right panel**. Frequency curve, Ω(r) calculated from the circular velocity plotted on the left panel. The bar length is marked as a solid vertical line and the critical radius defined as 1.4 times the bar length is indicated as a dotted vertical line, which separates between the fast rotator range (in green) from the slow rotator range (in blue). In red we mark two possible corotation radii and their corresponding pattern speed.

Angular velocity curves can also be used to depict the effect of the bar evolution on the corotation. To do so schematically we evaluate the bar at two different moments, $t_0$ and $t_1$ with $t_1>t_0$. The angular braking of the bar implies that it rotates more slowly while it is evolving, from $\Omega_{bar}^0$ at $t_0$ to $\Omega_{bar}^1$ at $t_1$, where $\Omega_{bar}^1 < \Omega_{bar}^0$, consequently the corotation is pushed farther out, from $r_{CR}^0$ to $r_{CR}^1 > r_{CR}^0$. This is illustrated in all the panels of Figure 15, where we plot the same angular velocity curve as in Figure 14, and in red we indicate the position of the bar pattern speed and the corotation radius at $t_0$ and $t_1$. Thus, the slowdown of the bar implies that the corotation occurs at larger radii, so the rate at which the corotation grows depends on the rate at which the bar angular rate decreases, which in turn depends on the angular momentum exchange rate with other structures of the galaxy (Winberg 1985; Athanassoula 2003), mainly with the dark matter halo. On the other hand, it is also known that the bar grows in size, and this does not, in general, occur at the same rate as the increase of the corotation radius. So, the evolution of the ratio of the corotation radius to the bar length depends on how fast/slow is the growth of the former compared to the latter. This leads to two possible scenarios; in the first one, the resonance radius grows faster than the size of the bar, so if the initial condition is that the bar is fast rotator under these conditions, if the bar will evolve to become slow rotator. However if the bar is initially a slow rotator, it will keep this status during the evolution. We can call these transitions as FS and SS, respectively. In the second scenario, the growth of the corotation radius is slower than that of the bar; in that case we can have SF and FF transitions depending on whether the bar is initially a slow or fast rotator, respectively. These four cases are schematically described in the four panels of Figure 15, where the bar size at $t_0$, $r_{bar}^0$ is marked as a solid vertical line, and the size of the bar at $t_1$ is indicated as a dashed vertical line. The green regions cover the fast rotator range for each corotation indicated in the panel; this means that if a bar ends in this green region, it is classified (conventionally) as fast rotator, otherwise it is a slow rotator. It is known from numerical simulations (since the early work of Sellwood, see for example Sellwood 1981, until recent studies such as Villa-Vargas et al. 2009, 2010; Martínez-Valpuesta et al. 2017) that the bar growth rate is often larger than the slowdown rate, and in that case the bar increases its size faster than its corotation is moving outwards. This is clearly illustrated in Figure 16, where we plot the time evolution of the bar length and the corotation radius for models 1 & 2, in the left and right panels, respectively. The green area shows the fast rotator region according to the conventional definition, so when the values of the bar length fall within the green region the bar is a fast rotator, otherwise it is a slow rotator. In the two models considered here, the bar is a slow rotator in the early stages, but it always ends as a conventionally fast rotator in spite of the braking of its angular rate. This implies that in the final stages of the bar evolution, when it has undergone considerable braking and is rotating slowly, that bar will be classified conventionally as a fast rotator. This is clearly in contradiction with the fact that bar has been substantially slowed down. Under these conditions, only the transitions SF and FF are possible (depicted in panels c and d of Figure 15, respectively). In the present study we find evidence of this contradiction in the terminology of bars in Figure 13, where the largest bars, which are most likely to be the most evolved bars, are rotating with the lowest angular velocities compared with the speed of the disk. This is in agreement with the predictions of the simulations, but all these bars would be conventionally classified misleadingly as fast rotators.

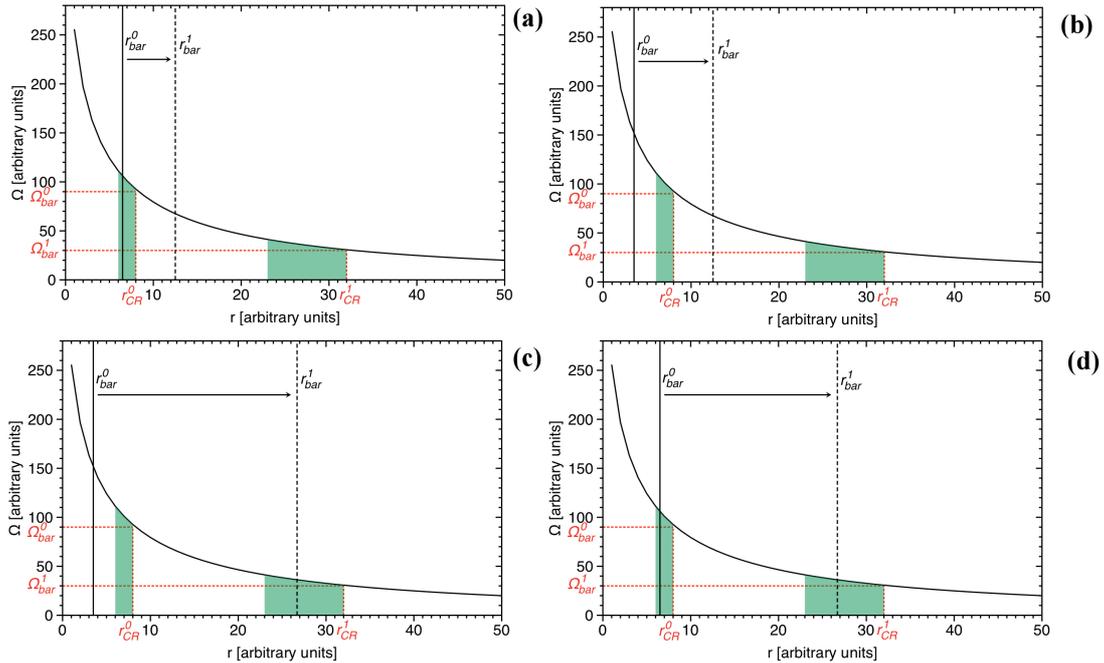

**Figure 15. Panel a.** The corotation radius and the bar pattern speed evaluated at two moments of the evolution ($r_{CR}^0$, $\Omega_{bar}^0$) at $t_0$, and ($r_{CR}^1$, $\Omega_{bar}^1$) at $t_1$, where $t_1 > t_0$. The green areas mark the range of the bar length for which the bar is fast rotator. The solid and dashed vertical lines mark the bar size at $t_0$ and $t_1$, respectively. This panel illustrates the transition FS (see text for details). **Panel b.** Same as panel a but displaying the transition SS. **Panel c.** Same as panel a but for the transition SF. **Panel d.** Same as panel a the transition FF is displayed.

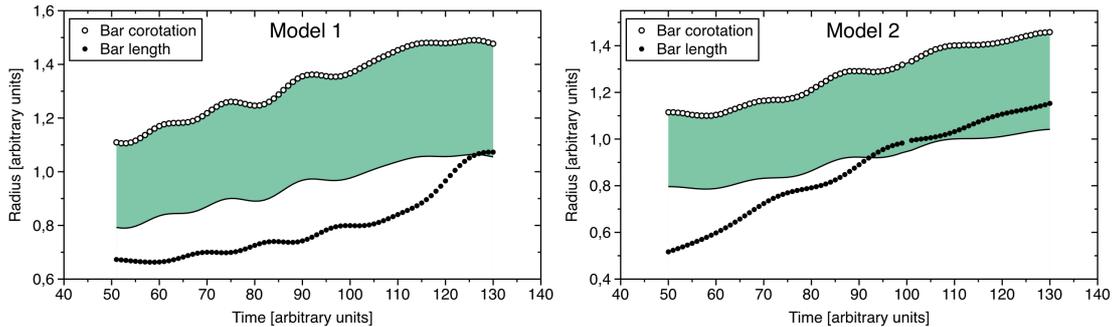

**Figure 16. Left panel.** Plot of the evolution of the bar corotation radius (open circles) and the bar length (black circles) for the simulations of model 1; the green region marks the fast rotator region, according to the conventional definition. **Right panel.** The same as in the left panel, but for model 2. We can see that for both models there are epochs when the bar has grown more quickly than the corotation radius.

## 6. CONCLUSIONS

In this study we have taken a sample of 68 barred galaxies observed with a Fabry-Pérot interferometer, for which we have applied the Font-Beckman method to determine the bar pattern speed and the bar corotation radius. We have also measured systematically the bar length and the bar strength from optical and near infrared images of the SDSS and the Spitzer surveys, respectively. The values of other parameters such as the total stellar mass and the 25-mag effective radius, $r_{25}$, of the galaxy are taken from databases. Collating this information on bars and galaxies we have analyzed how these parameters are related, and in some cases the results from numerical simulations of bar formation and evolution are used to provide a possible and plausible broad interpretation of the observational data. The conclusions we have reached can be summarized as follows:

- The distributions of the bar corotation to the bar length ratio shows 4 peaks centered at 1.09, 1.32, 1.59 and 1.85 (Figure 3), the first three are in good agreement with the values of this parameter found for the early, intermediate and late type galaxies, whilst the last one at 1.85 could correspond to galaxies that have experienced an interaction as numerical simulations of bar evolution predict higher values of the rotational parameter for these galaxies compared with those for non-interacting galaxies. When we plot the histogram of the bar pattern speed rationed with that of the outer disk, we find two populations, separated at the value 4 for this ratio.
- In Figure 4, we show that the bar corotation to the bar length ratio increases from morphological early type disk galaxies to intermediate types, and is then nearly constant up to the late types. The mean value of this dimensionless parameter for our sample is $\bar{\mathcal{R}} = 1.41 \pm 0.26$.
- Only intermediate and late type galaxies (which includes b, bc, c and d morphological types), show a uniform value of the mean scaled bar angular speed above 2.5, earlier and later types show lower values.
- A correlation between the bar length and the total stellar mass of the hosting galaxy is found (Figure 5, panel a). We do not see any relationship between the stellar mass and the corotation to the bar length ratio. Galaxies with intermediate stellar mass, $\log(M_*/M_\odot) \simeq 10$, harbor the fastest rotating bars and also the strongest bars (see Figure 5).

- In general, there is no relationship between the bar pattern speed and the stellar mass of the galaxy as is shown in **Figure 7**, however sorting the data according to the bar length, a good correlation between these magnitudes is found, but for only the shortest bars (with a size below 3 kpc).
- The largest bars rotate only with low pattern speeds and are hosted by the most massive galaxies. On the other hand, the fastest bars can only be small (with a length lower than 3 kpc). Finally, the least massive galaxies can only harbor short bars that rotate slowly (with an angular rate below ~40 km·s$^{-1}$·kpc$^{-1}$). Numerical simulations provide a useful interpretation of data plotted in Figure 7, in terms of bar slowdown and bar growth, as well as galactic mass growth, helping us to identify the least and the most evolved galaxies.
- Apparently there is no relation between the bar strength and the relative pattern speed, however we find that the strongest bars can rotate only very slowly compared with the velocity of the disk. We also show that those bars that rotate much more rapidly than the disk are always less strong. A strong anticorrelation between bar strength and bar pattern speed is obtained from the numerical simulations. This helps to define a preferred direction of evolution in the different diagrams where we interpret the data. Varying the distribution of the mass in the disk in the numerical simulations, it is possible to qualitatively cover the region in the plane ($S_b,\Gamma$) over which our cloud of data is extended. In this way it is possible to differentiate between those galaxies with a low contribution of dark matter in the disk, and those galaxies with a large fraction of dark matter in the disk.
- We have found that the bar strength is weakly correlated with the relative bar length, so the strongest bars tend to be the longest ones.

We show how confusing is the definition of fast/slow rotator bars using the criterion that the bar corotation radius to the bar length ratio is more/less than 1.4 In our sample there are many examples of galaxies that would be classified as fast rotators, which are rotating much more slowly than other bars conventionally classified as slow rotators. This classification is also in contradiction with a number of recent detailed simulation studies of bar evolution, according to which, the bar experiences a growth in size while its pattern speed declines. This can result in a very large bar relative to the size of the galaxy, which is moving with very low angular velocity, but would still be classified as a fast rotator because the ratio of its corotation radius to the bar length is well below 1.4. One consequence of the conventional classification has been that the observed effect of dark halos in braking bar rotation has been consistently underestimated. We propose the use of the bar pattern speed relative to the angular velocity of the disk, rather than the ratio between the corotation radius and the bar length, as a more suitable parameter to describe fast/slow rotator bars.

Acknowledgements

We would like to express our thanks to the anonymous referee whose comments have enabled us to make important improvements in the article. The research was carried out with funding from project P/308603 of the Instituto de Astrofísica de Canarias. The William Herschel Telescope is operated on the island of La Palma by the Isaac Newton Group in the Spanish Observatorio del Roque de los Muchachos of the Instituto de Astrofísica de Canarias. This research has made use of the Fabry Pérot database, operated at CeSAM/LAM, Marseille, France. This work is based in part on observations made with the Spitzer Space Telescope, which is operated by the Jet Propulsion Laboratory, California Institute of Technology under a contract with NASA. Funding for SDSS-III has been provided by the Alfred P. Sloan Foundation, the Participating Institutions, the National Science Foundation, and the U.S. Department of Energy Office of Science. The SDSS-III web site is http://www.sdss3.org/. SDSS-III is managed by the Astrophysical Research Consortium for the Participating Institutions of the SDSS-III Collaboration including the University of Arizona, the Brazilian Participation Group, Brookhaven National Laboratory, Carnegie Mellon University, University of Florida, the French Participation Group, the German Participation Group, Harvard University, the Instituto de Astrofisica de Canarias, the Michigan State/Notre Dame/JINA Participation Group, Johns Hopkins University, Lawrence Berkeley National Laboratory, Max Planck Institute for Astrophysics, Max Planck Institute for Extraterrestrial Physics, New Mexico State University, New York University, Ohio State University, Pennsylvania State University, University of Portsmouth, Princeton University, the Spanish Participation Group, University of Tokyo, University of Utah, Vanderbilt University, University of Virginia, University of Washington, and Yale University. We thank Peter Erwin for help in defining and measuring the lengths of galactic bars.

Appendix A. HOW TO ESTIMATE THE BAR PATTERN SPEED WITH A NEAR INFRARED IMAGE.

In Section 4, we investigate how the different magnitudes we have measured are related, and we have found a tight relationship between the bar pattern speed and its size (see figure 6). In panel a of Figure 17, we reproduce the same plot, but the data, now, are colored according to the morphological type of the hosting galaxy, the early types, i.e. galaxies of types {a,ab,b}, in red, the intermediate those galaxies of types{bc,c,cd}, in green, and the late type galaxies, those of type {d,dm,m}, in blue. We can see that early type galaxies are spread along the upper envelope of the cloud of points in the plot, and the late type of galaxies fall along the lower envelope of the cloud. So, in Figure 17, panels b, c and d, we plot (in log-log) the bar angular rate versus its length for the early, intermediate and late type of, respectively. A linear fit is performed for each morphological type of galaxy, as a there is a correlation between these two magnitudes. The results of the fit for each type of galaxy are:

Early type  $\Omega_{bar}(km \cdot s^{-1} \cdot kpc^{-1}) = 2.01 \cdot [r_{bar}(kpc)]^{-0.68}$  (A1)
Intermediate type  $\Omega_{bar}(km \cdot s^{-1} \cdot kpc^{-1}) = 1.78 \cdot [r_{bar}(kpc)]^{-0.51}$  (A2)
Late type  $\Omega_{bar}(km \cdot s^{-1} \cdot kpc^{-1}) = 1.53 \cdot [r_{bar}(kpc)]^{-0.42}$  (A3)

The correlation between the bar size and the bar angular rate can be used to estimate the latter given the former. So, with a near-infrared image (or alternatively, an image from SLOAN in r-band), we can measure the bar length by ellipse fitting, and knowing the morphological type of the galaxy, we apply the corresponding relationship to obtain an estimation of the bar pattern speed.

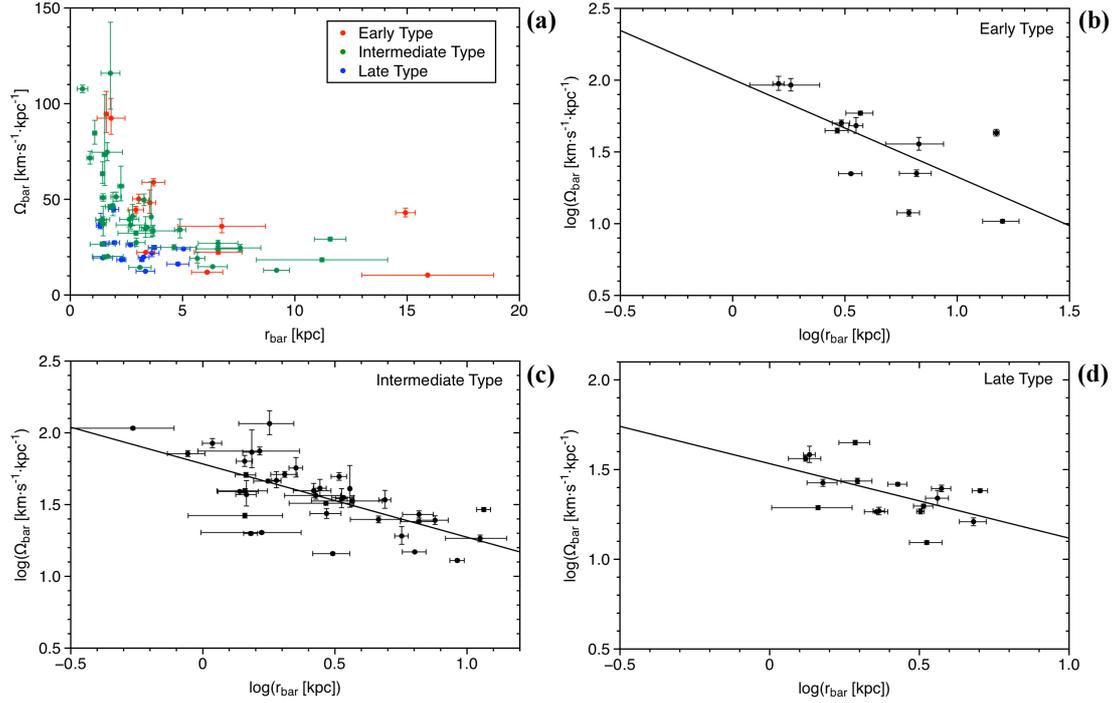

**Figure 17. Panel a.** Plot of the bar angular velocity versus the bar length. The color scale responds to the morphological type; Early type includes galaxies of type {a,ab,b}, intermediate does it with {bc,c,cd} and late type with {d,dm,m}. **Panel b.** Same as panel a, but now in log-log scale and only for the galaxies of early type. The solid line shows the linear fit to the data. **Panel c & d.** Same as panel b, but for the intermediate and late type of galaxies, respectively.